\documentclass[%
 reprint,
superscriptaddress,
 amsmath,amssymb,
 aps,
 pra,
]{revtex4-1}

\usepackage{graphicx}
\usepackage{dcolumn}
\usepackage{bm}
\usepackage{braket}
\usepackage{tabularx}
\usepackage[table]{xcolor}
\usepackage{threeparttable}
\usepackage{graphicx}
\usepackage{mathtools}
\usepackage{multirow}
\usepackage[colorlinks=true,linkcolor=blue,urlcolor=black,citecolor=blue,bookmarksopen=true]{hyperref}
\usepackage[caption=false]{subfig}
\usepackage{physics}
\usepackage{mathrsfs}
\usepackage{xifthen}
\usepackage{array}
\usepackage{enumitem}

\newcommand{\pnl}{Physical and Computational Sciences, Pacific Northwest National Laboratory, Richland, WA, 99354, USA}
\newcommand{\uw}{Department of Electrical and Computer Engineering, University of Washington, Seattle, WA, 98195, USA}
\newcommand{\ubc}{Department of Electrical and Computer Engineering, the University of British Columbia, Vancouver, BC V6T 1Z4, Canada}

\begin{document}

\preprint{APS/123-QED}

\title{Exact and Efficient Stabilizer Simulation of Thermal-Relaxation Noise for Quantum Error Correction}

\author{Sean R. Garner}
\email{Email: sean.garner@pnnl.gov}
\affiliation{\pnl}
\affiliation{\uw}

\author{Nathan M. Myers}
\affiliation{\pnl}

\author{Meng Wang}
\affiliation{\pnl}
\affiliation{\ubc}

\author{Samuel Stein}
\affiliation{\pnl}

\author{Chenxu Liu}
\email{Email: chenxu.liu@pnnl.gov}
\affiliation{\pnl}

\author{Ang Li}
\affiliation{\pnl}
\affiliation{\uw}

\date{\today}% It is always \today, today,

\begin{abstract}

Stabilizer-based simulation of quantum error-correcting codes typically relies on the Pauli-twirling approximation (PTA) to render non-Clifford noise classically tractable, but PTA can distort the behavior of physically relevant channels such as thermal relaxation. Physically accurate noise simulation is needed to train decoders and understand the noise suppression capabilities of quantum error correction codes. In this work, we develop an exact and stabilizer-compatible model of qubit thermal relaxation noise and show that the combined amplitude damping and dephasing channel admits a fully positive probability decomposition into Clifford operations and reset whenever $T_2 \leqslant T_1$. For $T_2 > T_1$, the resulting decomposition is negative, but allows a smaller sampling overhead versus independent channels. We further introduce an approximated error channel with reset that removes the negativity of the decomposition while achieving higher channel fidelity to the true thermal relaxation than PTA, and extend our construction to finite temperature relaxation. We apply the exact combined model to investigate large surface codes and bivariate bicycle codes on superconducting platforms with realistic thermal relaxation error. The differing logical performances across code states further indicate that noise-model-informed decoders will be essential for accurately capturing thermal-noise structure in future fault-tolerant architectures.

\end{abstract}

\maketitle

\section{Introduction}

% Quantum computing promises substantial computational speedup for problems that are classically hard, including integer factoring~\cite{}, sorting~\cite{}, combinatorial optimization~\cite{}, and quantum simulation of molecular and material systems~\cite{}. However, the practical implementation of large-scale quantum computation requires managing quantum noise and errors that originate from decoherence, control imperfections, and environmental coupling carefully. Quantum error correction (QEC) provides a systematic framework for suppressing these effects by redundantly encoding logical qubits into a subspace of many physical qubits' Hilbert space, where the information is protected by symmetries~\cite{}. Through repeated syndrome extraction and recovery operations, QEC can suppress the error rates exponentially with respect to the code distance, which forms the foundation for fault-tolerant quantum computing (FTQC)~\cite{}.

Careful suppression and managing the noise and decoherence are essential for reliable quantum computation, making quantum error correction (QEC) a central focus towards building large-scale fault-tolerant quantum computers~\cite{preskill_fault-tolerant_1997, preskill_quantum_2018, katabarwa_early_2024}. Recently, substantial progress has been achieved in both the theoretical design and experimental implementation of QEC across multiple physical platforms, including superconducting circuits~\cite{bravyi_future_2022, chou_superconducting_2024,  acharya_quantum_2025, lacroix_scaling_2025}, trapped ions~\cite{paetznick_demonstration_2024, reichardt_demonstration_2024}, and neutral atoms~\cite{sahay_high-threshold_2023, xu_constant-overhead_2024, bluvstein_fault-tolerant_2025, chiu_continuous_2025}. Advances in code constructions, such as surface codes~\cite{fowler_surface_2012,yoder_surface_2017,gidney_yoked_2025}, color codes~\cite{bombin_exact_2007, fowler_2-d_2011,kubica_unfolding_2015}, and quantum low-density parity check (qLDPC) codes~\cite{bravyi_high-threshold_2024, yoder_tour_2025, ruiz_ldpc-cat_2025}, have enabled experimental demonstrations of multi-round error correction and logical qubit surpassing the breakeven point and showing below-threshold performance. A central challenge that persists, however, is the accurate characterization of logical-level behavior under realistic noise. Thermal noise is especially pervasive throughout all quantum computers, yet has only thus far been captured by approximation or exponential exact simulation, as the relaxation channel is beyond the constraints of the Gottesman-Knill theorem~\cite{gottesman_heisenberg_1998}.

Reliable evaluation of code performance, threshold estimation, and decoder benchmarking all depend on understanding how physical noise models propagate through the error-correction process. To better access candidate QEC codes and architectures operating in realistic experimental environments, we need efficient simulation tools that can incorporate more realistic noise models and scale to thousands of physical qubits. 

Conventional full-state noise simulation approaches, such as state-vector (SV) and density-matrix (DM) methods, represent arbitrary quantum processes by explicitly evolving the quantum states. The universality allows them to capture any completely positive trace-preserving (CPTP) map acting on the system, but at the cost of exponential scaling with $n$ number of qubits ($O(2^n)$ and $O(4^n)$ for SV and DM respectively). This makes them intractable for moderate-sized QEC codes. Previous efforts investigated the QEC code performance under realistic noise using DM simulations, however, they are limited to small code distances and rely on more simplifications and approximations, such as re-using the syndrome qubits in the simulation, which potentially violates the temporal order of syndrome measurements in experimental implementations~\cite{katabarwa_logical_2015, geller_efficient_2013}. These challenges make the stabilizer simulation framework necessary for simulating QEC code performance, at the cost of exactness~\cite{aaronson_improved_2004}.

% These challenges make the stabilizer simulation framework, which efficiently tracks Clifford gates and Pauli-basis measurements using polynomial classical resources, promising for simulating QEC code performance~\cite{aaronson_improved_2008, gidney_stim_2021}.

The stabilizer formalism allows Clifford circuits to be simulated in $O(n^2)$ time and memory by evolving a tableau that tracks the stabilizers' transformation under Clifford operations~\cite{gottesman_stabilizer_1997}. This efficiency makes it widely adopted in most QEC simulation tools, such as Stim~\cite{gidney_stim_2021}, and enables exploration of mid- to large-scale QEC code patches in the near FTQC regime. However, the efficiency of stabilizer simulation relies on the restriction that the operations are within the Clifford group, with noise channels that are expressed as stochastic Pauli channels. In practice, non-Clifford physical error channels, such as coherent errors~\cite{bravyi_correcting_2018, huang_performance_2019} and thermal errors~\cite{geller_efficient_2013, georgopoulos_modelling_2021}, cannot be directly simulated in stabilizer simulators. Instead, the Pauli-twirling approximation (PTA) is commonly employed to replace a general CPTP map by its depolarized Pauli channels~\cite{divincenzo_quantum_2002, dur_standard_2005, ghosh_surface_2012}. Although efficiently simulable, previous studies on small codes have shown that PTA can underestimate performance, producing logical error rates that are several times larger than true values~\cite{tomita_low-distance_2014,katabarwa_logical_2015}.

A more accurate yet still Clifford-compatible approach is the quasi-probabilistic decomposition (QPD) of a non-Clifford channel into a linear combination of Clifford channels~\cite{bennink_unbiased_2017}. By sampling from a rescaled non-negative probability distribution and making a weight adjustment, one can exactly reproduce the original channel statistics with only Clifford simulators. However, negativity in the probabilistic distribution results in an exponential sampling cost, i.e., the estimator variance will grow exponentially with the number of non-Clifford error channels~\cite{bennink_unbiased_2017,katsuda_simulation_2024}. To model qubits' thermal relaxation errors in the QEC circuits, the cost will grow with circuit depth and number of qubits. Consequently, while QPD yields unbiased results, it can introduce exponential sampling cost, hindering its implementation for efficiently simulating realistic error models for large-scale QEC code performances. Moreover, QPD often requires explicit stabilizer-state reset operations in some sampling branches of the error channel, which are not directly compatible with Pauli-frame tracking for stochastic Pauli error channels~\cite{gidney_stim_2021}, further slowing down the simulation.

% \textcolor{red}{AL: need to very briefly indicate that: (i) relaxation error channel is not Clifford simulatable; (ii) relaxation in comparison is more realistic modeling of the noise and has been adopted by existing works.}\sg{(i) is now adressed in paragraphs 1 and 2. (ii) compare existing works with small releastic noise modeling here}
In this paper, we focus on the qubit thermal relaxation error channel, which is one of the most commonly seen non-Clifford error sources in experiments. We show that combining the amplitude damping ($T_1$) and dephasing ($T_2$) channels into a unified relaxation channel significantly reduces the QPD sampling overhead required for exact Clifford-based simulation. Specifically, we find that when  $T_2 \leq T_1$, a regime typical for superconducting qubits and several solid-state qubit platforms~\cite{acharya_quantum_2025, abughanem_ibm_2025}, the thermal relaxation channel admits a fully positive decomposition into Clifford channels. In the regime $T_2 > T_1$, we demonstrate that the resulting decomposition is still less costly to sample than treating amplitude and phase damping separately. We further introduce a reset-based approximation that can preserve complete positivity of the decomposition, while achieving substantially higher channel fidelity than the PTA model. We incorporate these constructions into an MPI-accelerated and GPU-capable stabilizer simulator and evaluate the logical performance of both surface codes and bivariate bicycle (BB) codes under realistic thermal-relaxation noise, beyond the scope of previous works that only focus on realistic noise performance on surface code only or other small QEC codes~\cite{tomita_low-distance_2014, puzzuoli_tractable_2014, Gutierrez2015, bravyi_correcting_2018, Marton2023coherenterrors, Darmawan2017, Schwartzman-Nowik2025}. Our simulations show that PTA can misestimate logical error rates by several factors, either overestimating or underestimating, depending on code distance and logical basis, whereas the composite and reset-based treatments accurately capture the directional bias of relaxation. We also observe that logical error rates depend on the encoded logical state, suggesting that bias-aware, noise-informed decoding strategies are needed for future fault-tolerant architectures.

% To exploit this for large-scale code analysis, we demonstrate a GPU-accelerated stabilizer simulator that utilizes parallelization to efficiently simulate Clifford operations. Using this simulator, we investigate the logical performance of the surface code and the bivariate bicycle (BB) code under exact $T_1$ and $T_2$ thermal noise, and compare them against the corresponding channels under the PTA. We find that a PTA isolating just damping and decoherence can both underestimate and overestimate logical errors by an order of magnitude versus a true $T_1$ and $T_2$ noise model, depending on code construction, and can affect the rate of error suppression entirely. For BB $\ket{0}_L^{\otimes 12}$ state memories and $\ket{+}_L$ surface code, $\ket{0}$ population bias not captured in the PTA is amortized such that the PTA is a near constant overestimation of logical error. In small, near-term $\ket{0}_L$ surface code memories, this bias impacts the rate of error suppression when unaccounted for in the decoder.
% ... (for Sean: add one sentence to summarize the main finding you got from this work.)

Our paper is organized as follows. Sec.~\ref{sec:theory} reviews the theoretical foundations of thermal relaxation error models, including their Kraus representations, the connection to relaxation dynamics induced by a thermal bath, the Pauli-twirling approximation (PTA), and the exact quasi-probabilistic decomposition of amplitude-damping processes. We then present our treatment of the thermal relaxation channel, identify the parameter regimes in which it admits exact Clifford-compatible simulation, and introduce an improved approximation in regimes where this exactness does not hold. Sec.~\ref{sec:sim} applies the resulting error model to evaluate the performance of the surface code and the BB code on superconducting-qubit platforms. We provide further discussion in Sec.~\ref{sec:discussion} and conclude in Sec.~\ref{sec:summary}.

\section{Theoretical Modeling of Thermal Relaxation Errors} \label{sec:theory}

Thermal relaxation is a fundamental and widely encountered noise mechanism in quantum computing. Because the exact thermal relaxation channel is inherently non-Pauli and cannot be represented as a stochastic Pauli channel, approximations are required to incorporate it into the current stabilizer-based QEC simulation toolchain. 

In this section, we review the main theoretical treatment of thermal relaxation and introduce the model adopted in this work. Sec.~\ref{sec:theory:thermal} provides a brief review of the thermal relaxation channel. Sec.~\ref{sec:theory:PTA} reviews the Pauli twirling approximation and its application to thermal relaxation, highlighting its limitations. Sec.~\ref{sec:theory:QPD} reviews prior study on quasi-probabilistic decomposition (QPD) and its application to the amplitude damping channel. In Sec.~\ref{sec:theory:QPD_thermal}, we extend the QPD to general thermal relaxation error channels, identifying the efficiency gain when combining the transverse and longitudinal relaxation processes and treating them jointly. We then provide an improved approximation of the thermal relaxation in the relevant parameter regime with better accuracy, which is well-suited for stabilizer-based QEC simulations.

\subsection{Thermal Relaxation Error Channels} \label{sec:theory:thermal}
% goal: intro thermal channel with Temperature 0 and nonzero temperature
% intro the Kruas operator representations
% link Kraus channels with the T1 T2 time,
% intro necessary assumptions, notations and conventions

Quantum computing systems are inevitably coupled to their environment, which causes qubits to thermalize towards the bath environment. Thermal relaxation can lead to energy loss, causing qubits to relax from the excited state to the ground state and lose the coherence between the computational states. The relaxation effects are characterized by the qubit energy relaxation time $T_1$ and the decoherence time $T_2$. 

To model the effect of thermal noise on the quantum systems, the relaxation dynamics can be described by a master equation and integrated for a time period $\tau$. For a single qubit interacting with a thermal bath, the master equation is
\begin{align}
    \partial_t \rho & = \gamma (\langle n_b\rangle +1) \mathcal{D}[\dyad{0}{1}](\rho) + \gamma \langle n_b \rangle \mathcal{D}[\dyad{1}{0}](\rho) \nonumber \\ 
    & + \frac{\gamma_\phi}{2} \mathcal{D}[\sigma_z](\rho),
    \label{eq:master}
\end{align}
where relaxation rate $\gamma$ is determined by the qubit-bath coupling, $\langle n_b \rangle$ is the thermal occupation of the bath at temperature ($T_b$), given by $\langle n_b \rangle = 1 / (e^{\hbar \omega/k_b T_b} - 1)$, and $\Gamma_\phi$ characterizes the extra dephasing effect on the qubit. The dissipator in Eq.~\eqref{eq:master} is
\begin{equation}
    \mathcal{D}[\hat{A}](\rho) = - \frac{1}{2} \left( \hat{A}^\dagger \hat{A} \rho + \rho \hat{A}^\dagger \hat{A} - 2 \hat{A} \rho \hat{A}^\dagger\right),
\end{equation}
for any jump operator $\hat{A}$ and system density operator $\rho$~\cite{Scully1997, Gardiner2004quantum}. 

When the qubit couples to a zero-temperature bath ($\langle n_b \rangle = 0$) for a time $\tau$, the density operator evolves to
\begin{equation}
\rho(\tau) =\begin{pmatrix}
1-\rho_{11} e^{-\tau/T_1} & \rho_{01} e^{-\tau/T_2} \\[2pt] 
\rho_{10} e^{-\tau/T_2} & \rho_{11} e^{-\tau/T_1} \end{pmatrix},
\label{eq:thermal_evolve}
\end{equation}
where $\rho_{i,j} = \matrixel{i}{\rho(t=0)}{j}$, $T_1=1/\gamma$, and $T_2 = 1/(\gamma/2 + \Gamma_\phi)$. As the dephasing between the state $\ket{0}$ and $\ket{1}$ comes from two sources: the energy relaxation and pure dephasing effects, the decoherence time satisfies 
\begin{equation}
   \frac{1}{T_2}=\frac{1}{2T_1}+\frac{1}{T_\phi},
   \label{eq:t2}
\end{equation}
where $T_\phi = 1/\gamma_\phi$ is the pure dephasing time. From Eq.~\eqref{eq:t2}, the qubit dephasing time $T_2$ is bounded by $2T_1$.

The thermal relaxation process can be equivalently described by Kraus operators~\cite{nielsen}. When there is no pure dephasing, $T_2 = 2 T_1$, the process reduces to the amplitude damping channel as $\mathcal{E}_{\text{ad}}(\rho) = E_0 \rho E_0^\dagger + E_1 \rho E_1^\dagger$, with
\begin{align}
    E_{\text{ad}, 0} &= \left( \begin{array}{cc}
        1 & 0 \\
        0 & \sqrt{1-p_\gamma}
    \end{array} \right), 
    \quad E_{\text{ad}, 1} = \left( \begin{array}{cc}
        0 & \sqrt{p_\gamma} \\
        0 & 0
    \end{array} \right),
    \label{eq:kraus_ad}
\end{align}
where the channel is characterized by a ``relaxation probability'' $p_\gamma$~\cite{nielsen}. Comparing with the density operator evolution in Eq.~\eqref{eq:thermal_evolve}, the relaxation probability is
\begin{equation}
    p_\gamma(\tau) = 1 - e^{-\tau / T_1}.
    \label{eq:pgamma}
\end{equation}

When the pure dephasing presents, which causes the decoherence time $T_2 < 2 T_1$, the additional dephasing noise can be modeled as an additional stochastic Pauli-Z error with probability $p_{\phi}$, represented in Kraus operators as
\begin{align}
    \mathcal{E}_{\text{pd}}(\rho) = (1 - p_\phi) \rho + p_\phi Z \rho Z.
    \label{eq:dephasing}
\end{align}
The full thermal relaxation process is
\begin{align}
    \mathcal{E}_{\text{pd}}[\mathcal{E}_\text{ad}(\rho)] \stackrel{\mathrm{def}}{=} (\mathcal{E}_{\text{pd}} \circ \mathcal{E}_\text{ad})(\rho).
\end{align}
Matching Eq.~\eqref{eq:thermal_evolve} yields the additional Pauli-Z error probability,
\begin{equation}
    p_\phi = \frac{1}{2} \left[ 
        1 - e^{-\tau \left(\frac{1}{T_2} - \frac{1}{2T_1}\right)}
    \right] = \frac{1}{2} \left( 1 - e^{-\tau / T_\phi} \right).
    \label{eq:pphi}
\end{equation}

At finite bath temperature, the bath thermal occupation $\langle n_b \rangle \neq 0$, with the value of $\langle n_b \rangle$ determined by the bath temperature $T_b$. Similarly, we can integrate the full master equation Eq.~\eqref{eq:master} to solve the dynamics of the qubit. Specifically, when $\tau \rightarrow \infty$, the population in the excited state is 
\begin{equation}
    p_1 = \frac{\langle n_b \rangle }{1 + 2 \langle n_b \rangle} = \frac{1}{1 + e^{\hbar \omega_0 / k_b T_b}}.
    \label{eq:p1}
\end{equation}
The qubit's total decay rate to the equilibrium state defines the qubit's $T_1$ time as $1/T_1 = \gamma (2\langle n \rangle +1)$, while the dephasing rate defines the decoherence time $T_2$ as $1/T_2 = \gamma(\langle n_b\rangle + 1/2) + \Gamma_\phi$.

% Similarly, the thermal relaxation process with finite temperature $T_b$ can be represented by Kraus operators. 
When there is no pure dephasing ($\Gamma_\phi = 0$), the finite-temperature thermal relaxation process can be described by a generalized amplitude damping channel as~\cite{nielsen},
\begin{equation}
    \mathcal{E}_{\text{gad}}(\rho) = \sum_{i=1}^{4} E_{\text{gad},i} \rho E_{\text{gad},i}^\dagger,
\end{equation}
with the Kraus operators,
\begin{align}
    E_{\text{gad}, 0,1} & = \sqrt{1-p_1} E_{\text{ad},0,1}, \label{eq:gad_kraus_01} \\
    E_{\text{gad}, 2} & = \sqrt{p_1} \left( 
    \begin{array}{cc}
      \sqrt{1-p_\gamma'}   & 0 \\
         0 & 1
    \end{array}\right), \, \label{eq:gad_kraus_2} \\
    E_{\text{gad}, 3} & = \sqrt{p_1} \left(
    \begin{array}{cc}
       0  & 0 \\
       \sqrt{p_\gamma'}  & 0
    \end{array}\right), \label{eq:gad_kraus_3}
\end{align}
where $p_\gamma$ defined in $E_{\text{ad}, 0,1}$ becomes 
\begin{equation}
    p_\gamma = p_\gamma' = 1- e^{- \gamma \tau (2 \langle n_b \rangle +1)} = 1 - e^{-\tau/T_1},\label{eq:pgamma_temp}
\end{equation}
$p_1$ is the population of $\ket{1}$ state when $\tau \rightarrow \infty$ as given in Eq.~\eqref{eq:p1}.

When there are additional dephasing ($\Gamma_\phi \neq 0$), the thermal relaxation process can be modeled as $\mathcal{E}_{\text{pd}} \circ \mathcal{E}_\text{gad}$, where the Pauli-Z error probability $p_\phi$ is given in Eq.~\eqref{eq:pphi}.
    
%, the qubit population can be thermally excited from the ground state to the excited state. 

\subsection{Pauli Twirling and Pauli Twirling Approximation to Quantum Channels} \label{sec:theory:PTA}

Twirling is a technique used to symmetrize the effect of a noise channel by randomly applying a unitary operation from a chosen symmetry group $\mathcal{G}$ and its inverse. This process effectively averages the noise channel over the group $\mathcal{G}$, which is widely used for both characterizing and mitigating noise in quantum systems~\cite{geller_efficient_2013,wallman_noise_2016}. Depending on the chosen $\mathcal{G}$, various twirling methods arise, such as unitary-twirling~\cite{stonner_deterministic_2020}, Clifford-twirling~\cite{tsubouchi_symmetric_2025}, and Pauli-twirling~\cite{geller_efficient_2013}. Among these, Pauli twirling is one of the most popular choices due to its relative simplicity in implementation and analysis. 

Pauli twirling involves averaging the channel over the $n$-qubit Pauli group $\mathcal{P}_n$, which transforms the original arbitrary noise channel into stochastic Pauli channels. Approximating the original channel by its Pauli-twirled version is known as the Pauli twirling approximation (PTA). For the original channel $\mathcal{E}$, the channel after Pauli twirling is
\begin{equation}
    \mathcal{E}_{\text{pta}}(\rho) = \sum_{\sigma_i \in \mathcal{P}_n} \sigma_i \mathcal{E}(\sigma_i \rho \sigma_i) \sigma_i,
\end{equation}
where $\mathcal{P}_n$ is the $n$-qubit Pauli group, $\sigma_i$ are the Pauli operators. We use the fact that the Pauli operators are unitary and Hermitian. 
Notice that a quantum channel can be equivalently described by its $\chi$ matrix as,
\begin{equation}
    \mathcal{E}(\rho) = \sum_{i,j} \chi_{i,j} \sigma_i \rho \sigma_j.
\end{equation}
It is shown that the $\chi$ matrix of the PTA channel is diagonal,
\begin{equation}
    \mathcal{E}_{\text{pt}}(\rho) = \sum_{i,i} \chi_{i,i} \sigma_i \rho \sigma_i,
\end{equation}
i.e., Pauli twirling removes all the off-diagonal elements in the original $\chi$ matrix and retains only the diagonal coefficients. 

The PTA replaces an arbitrary quantum channel with an incoherent mixture of Pauli error channels. It is widely used in analyzing the effects of complex noise models in quantum algorithms and quantum error correction, because the resulting channel is efficiently simulable using stabilizer simulators for large-scale physical qubits in QEC codes~\cite{katabarwa_logical_2015, tomita_low-distance_2014, geller_efficient_2013}. 

\begin{figure}[htbp]
    \centering
    \subfloat[]{\includegraphics[width=0.45 \linewidth]{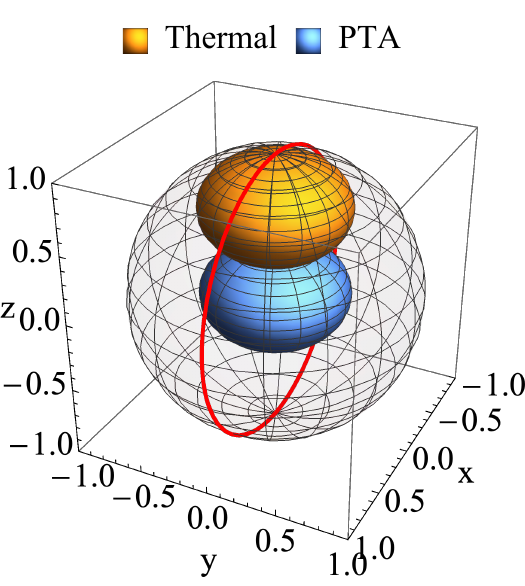}}
    \subfloat[]{
    \includegraphics[width=0.47 \linewidth]{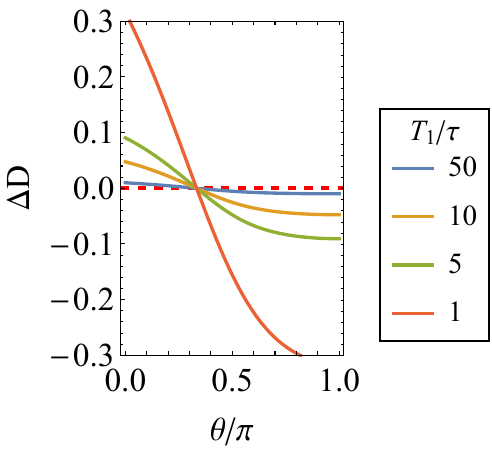}
    }
    \caption{The effect of the thermal relaxation error channel and the channel after PTA. (a) The Bloch sphere after applying the thermal relaxation channel and the channel after PTA on a single qubit. We notice that the PTA channel can nicely capture the distortion on the Pauli-X and Y directions, while the directional relaxation along the Pauli-Z axis is not. The gray-shaded sphere shows the unit Bloch sphere. (b) The trace distance difference ($\Delta D$) between the two channels. We focus on the states whose Bloch vectors lie along the red great circle in (a). Parameters if not specified: $\tau/T_1 = 1.0$, $T_2/T_1 = 1.5$, thermal temperature $T_b = 0$.}
    \label{fig:bloch}
\end{figure}

Although the PTA provides a way to efficiently investigate the effect of complex noise, the twirled channel may induce significant bias to the true noise channel. Considering the thermal relaxation error channel in Sec.~\ref{sec:theory:thermal} as an example, when the thermal temperature is zero, the channel after the PTA is
\begin{equation}
    \mathcal{E}_{\text{th,pta}}(\rho) = (1-\sum_i p_i) \rho + \sum_{i} p_i \sigma_i \rho \sigma_i,
    \label{eq:th0_pta}
\end{equation}
where the sum runs over $\{x, y, z\}$ and the error probabilities are
\begin{equation}
    p_x = p_y = \frac{p_\gamma}{4}, \quad p_z = \frac{1}{2} - \frac{p_\gamma}{4} - \frac{(1-2p_\phi) \sqrt{1-p_\gamma}}{2},
\end{equation}
where $p_\gamma$ and $p_\phi$ are defined in Eqs.~\eqref{eq:pgamma} and~\eqref{eq:pphi}, respectively~\cite{geller_efficient_2013}. Fig.~\ref{fig:bloch} compares the action of the true thermal relaxation error channel and its PTA counterpart on the Bloch sphere. The contraction along the $X$-$Y$ plane is well captured, but the PTA fails to reproduce the directional relaxation along the $Z$ axis. 
%, we plot the effect of the thermal relaxation error channel and the channel after PWA on a single-qubit Bloch sphere. The noise effect on the X and Y directions is nicely captured by the PWA channel, however, the bias on the Z axis is not. 

% As the $n$-qubit Pauli operators form an orthogonal basis for the operator space, any density matrix can be expanded to this basis. Therefore, the Pauli transfer matrix (PTM) of a channel $\mathcal{E}$, defined as
% \begin{equation}
    % R_{ij} = \frac{1}{2^n} \Tr[\sigma_j \mathcal{E}(\sigma_i)], 
% \end{equation}
% is an equivalent representation of the channel. 

As the PTA does not accurately capture the full structure of thermal relaxation, Fig.~\ref{fig:bloch}a illustrates that it can either overestimate or underestimate the true error depending on the input state. Consider two extreme input states, $\ket{0}$ and $\ket{1}$. At zero temperature, thermal relaxation leaves state $\ket{0}$ unchanged, while the PTA introduces a non-zero error probability, leading to overestimating the error. Conversely, as thermal relaxation drives state $\ket{1}$ to state $\ket{0}$, the PTA channel underestimates the error for state $\ket{1}$. To quantify this effect, we compute the trace distance between the noisy states produced by the PTA channel and the noiseless state, and then compare it against the corresponding trace distance under the true thermal relaxation channel
\begin{equation}
    \Delta D \equiv D(\rho, \mathcal{E}_{\text{th}}^\text{PTA}[\rho]) - D(\rho, \mathcal{E}_\text{th}[\rho]),
\end{equation}
where $D$ denotes the trace distance operation. We evaluate this trace distance difference for the pure states $\ket{\psi(\theta)} = \cos(\theta/2) \ket{0} + \sin(\theta/2)\ket{1}$, whose Bloch vectors lie on the great circle with $y=0$ (red curve in Fig.~\ref{fig:bloch}a).
Fig.~\ref{fig:bloch}b shows the resulting trace distance differences. When $\theta<\pi/3$ (closer to $\ket{0}$), the PTA channel overestimates the error, whereas for  $\theta > \pi/3$ (closer to $\ket{1}$), the PTA channel underestimates the error. These observations motivate the need for more accurate representations of error channels than those provided by the PTA.

\begin{figure*}[htb]
    \centering
    \subfloat[]{
    \includegraphics[width= 0.4 \linewidth]{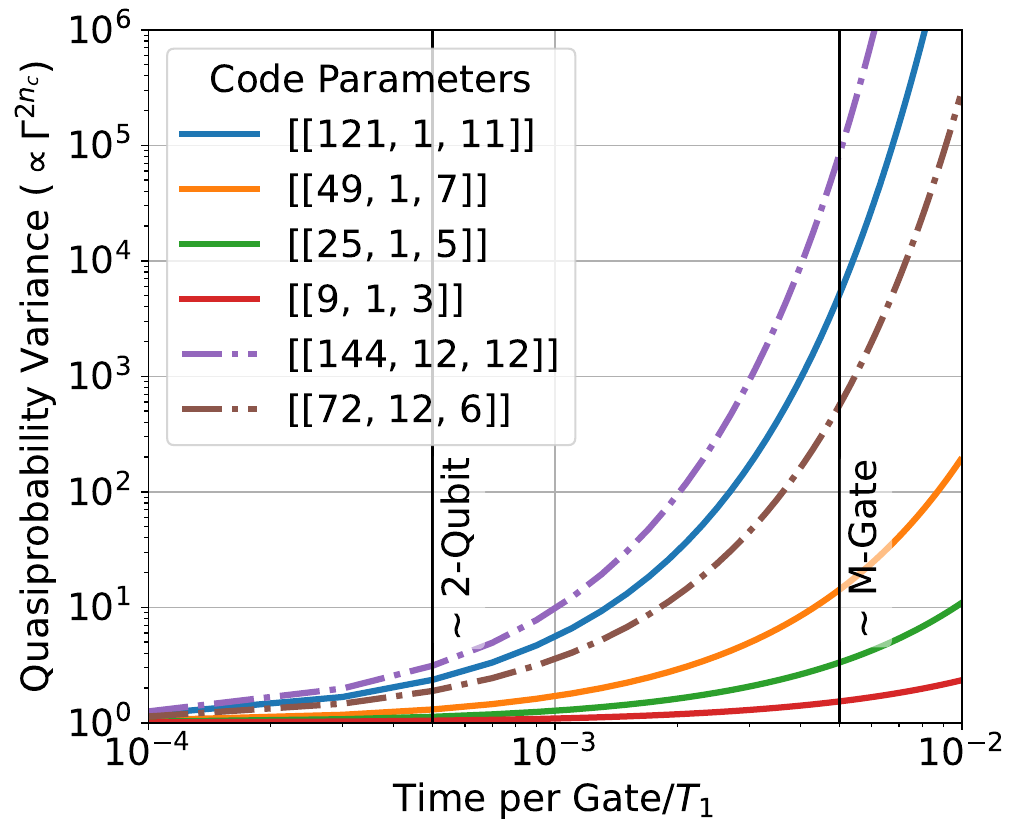}
    } \quad
    \subfloat[]{
    \includegraphics[width=0.4 \linewidth]{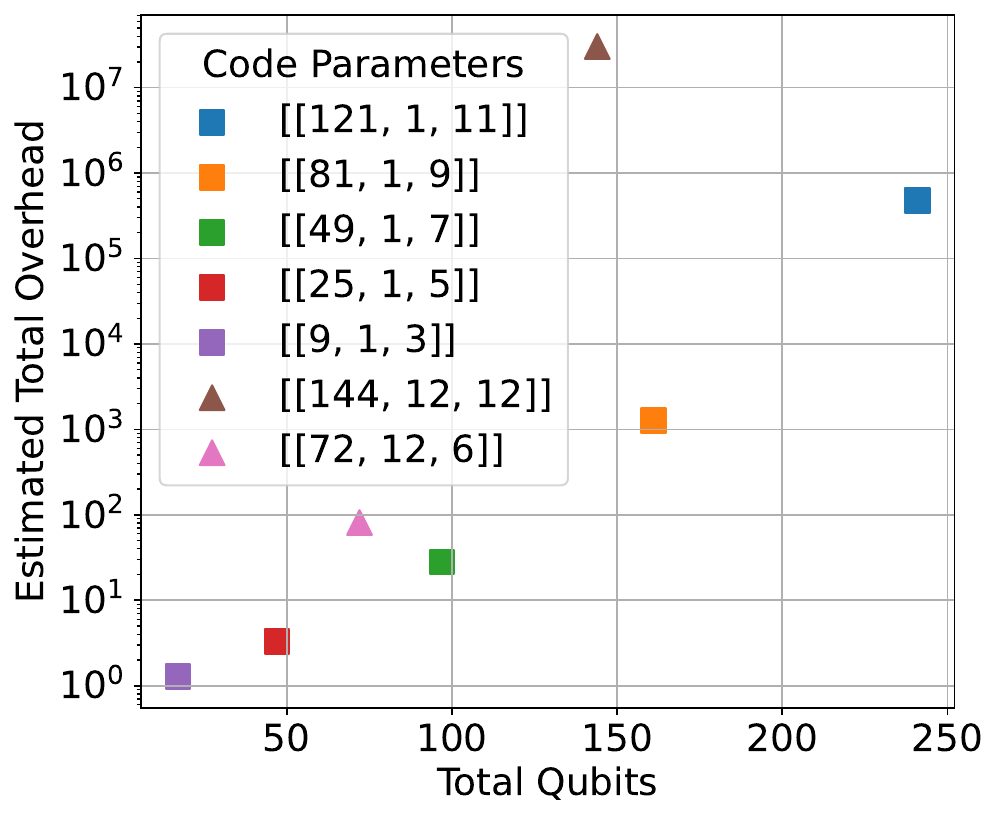}
    }
    \caption{(a) Sampling cost to offset estimator variance  $\propto\Gamma^{2 n_c}$ if all gates of a code have a uniform time cost, i.e., $\Gamma$ is the same for every type of gate. Vertical lines denote approximately the current time cost of measurement and 2-qubit gates in relation to $T_1$ coherence on IBM devices. In reality, the time cost will fall in between measurement (M) and 2-qubit gates, depending on the relative amount of those gates for each code. Overhead assumes $d$ rounds of syndrome extraction with a distance $d$ code. (b) Approximate quasi-probability sampling cost per-memory experiment. As opposed to (a), $\Gamma$ here is calculated for each gate time, for the number of each gate type in each code, representing the actual simulation. Measurement is approximated to take $T_1/100$, while 2-qubit gates take $T_1/1000$. The uncombined quasi-probability variance overhead is independent of $T_2$ as the probability decomposition for $T_2$ is always positive.}
    % approximated from IBMQ data including Table~\ref{table:IBM_data}.}
    \label{fig:sampling_cost_line}
\end{figure*}

\subsection{Quasi-probabilistic Decomposition (QPD) to Clifford Channels} \label{sec:theory:QPD}

Ref.~\cite{bennink_unbiased_2017} proposed a method to decompose an arbitrary quantum channel into a composition of Clifford and reset to stabilizer state operations. This approach begins by solving the Pauli-transfer matrix (PTM)~\cite{chow_universal_2012,hantzko_pauli_2025} of the quantum channel, and then solving for a decomposition into Clifford and reset operations element-by-element. As a concrete example, the amplitude damping channel in Eq.~\eqref{eq:kraus_ad} can be decomposed as
\begin{align}
    \mathcal{E}_{\text{ad}} = & q^{\text{(ad)}}_{+}\mathcal{I}+q^{\text{(ad)}}_{-} \mathcal{Z} + p_\gamma \mathcal{R}_{\ket{0}},
\label{eq:amp_decomp}
\end{align}
where $\mathcal{I}$, $\mathcal{Z}$ are identity and Pauli-Z channels, $\mathcal{R}_{\ket{0}}$ is the channel corresponding to reset to state $\ket{0}$. The quasi-probabilities are 
\begin{equation}
    q^{\text{(ad)}}_{\pm} = \frac{1-p_\gamma \pm \sqrt{1-p_\gamma}}{2}.
\end{equation}
This decomposition enables Monte-Carlo simulation of $T_1$ relaxation in QEC codes within the stabilizer-based framework. 
%can be used in simulating the $T_1$ relaxation in QEC codes, which can be fitted into stabilizer-based quantum simulation frameworks.

From Gottesman-Knill theorem~\cite{gottesman_stabilizer_1997}, it is unlikely that any arbitrary quantum channel can be classically simulated in an efficient manner. The cost of the above method is that probabilistic decomposition does not necessarily yield a completely positive distribution. Eq.~\eqref{eq:amp_decomp} shows the coefficient of the Pauli-Z channel to be always negative, meaning the decomposition yields a quasi-probabilistic (QP) distribution instead.

Although the resulting QP distribution is not a true probability distribution, it can still be sampled to estimate statistical expectations of any target observable, at the expense of increased sampling variance. For a random variable $x$, the expectation value of a function $f(x)$ on a QP distribution $q(x)$ is
\begin{equation}
    \langle f \rangle_q = \sum_x q(x) f(x).
\end{equation}
To sample from this QP distribution, a true probability distribution is defined as,
\begin{equation}
    p(x) = q(x) / \Gamma, \quad \Gamma = \sum_x \left\vert q(x) \right\vert,
\end{equation}
where $\Gamma$  measures the negativity of the QP distribution. The expectation value can then be rewritten as
\begin{equation}
    \langle f \rangle_q = \sum_x  \Gamma \operatorname{sgn}[q(x)]p(x)f(x)_q,
\label{eq:unbiased_estimator}
\end{equation}
which can be sampled using Monte-Carlo methods, resulting in an unbiased estimator for $\langle f \rangle$.

However, when sampling the QP distribution using Monte-Carlo methods, the variance of the estimator will grow with the negativity of the QP distribution~\cite{bennink_unbiased_2017,hakkaku_sampling-based_2021, piveteau_quasiprobability_2022}, which requires more Monte-Carlo shots to achieve a target precision. For the amplitude damping channel, the negativity is controlled by the ratio $\tau/T_1$. When there are multiple quantum channels in a quantum circuit that need to be decomposed independently, the total negativity causes the variance to multiply across each quantum channel, 
%when sampling the whole quantum circuit, the negativity of the total QP distribution is
\begin{equation}
    \Gamma_\text{total} = \prod_{i =1}^{n_c} \Gamma_i,
\end{equation}
where $\Gamma_i$ represents the negativity of each QP distribution and $n_c$ is the total number of decomposed channels. Consequently, the sampling cost must grow exponentially with both the number of qubits and the number of amplitude damping error channels (the depth of the quantum circuit), to offset the variance introduced by the unbiased estimator. Fig.~\ref{fig:sampling_cost_line} shows the total variance of the uncombined QP distributions when simulating different QEC codes and counting for all the amplitude damping error channels on each of the operations. In Fig.~\ref{fig:sampling_cost_line}a, we observe that as the gate time increases (larger error rates in each amplitude damping channel), the sampling overhead grows rapidly with the error strength. Furthermore, in Fig.~\ref{fig:sampling_cost_line}b,  we show the total variance as a function of the number of total qubits in each code experiment. It is clear that the sampling overhead can make numerical simulation of large-distance codes impractical when trying to determine logical error performance.

% The sampling cost to achieve some level of precision by Monte-Carlo sampling a single instance of a quasi-probability distribution versus sampling on a completely positive distribution can be quantified in terms of $\Gamma$. For $d$ channels in the simulation circuit, each with a uniform relaxation probability, 
% \begin{equation}
%     \Gamma_{total} = \prod_d \Gamma
% \end{equation}
% The variance of an estimator for the exact channel $\widehat f$ is defined by $\operatorname{Var} (\widehat f) = \langle (\widehat f  - \langle \widehat f
% \rangle)^2 \rangle$.
%  Substituting out $\Gamma^d$ from $\widehat f$, we find $\operatorname{Var}(\widehat f)\propto \Gamma^{2d}$, whose scaling agrees with the analysis in~\cite{bennink_unbiased_2017}.

\subsection{QPD on the Thermal Relaxation Channel} \label{sec:theory:QPD_thermal}

With the understanding that non-Clifford channels can be decomposed into stabilizer channels using QPD, and the cost to do so, it is natural to seek ways to reduce the negativity by minimizing $\Gamma$ from the QP distribution or the number of channels $n_c$. 

% Hakkaku {\it et al.}~\cite{hakkaku_sampling-based_2021} quantify the difficulty of a stabilizer decomposition in terms of channel robustness~\cite{seddon_quantifying_2019}, which is a measure of channel magic. They demonstrated that multiple coherent errors can result in a lower channel robustness and improve the simulation efficiency, enabling surface-code simulations up to distance-7 with coherent rotation errors.

As thermal relaxation is among the most common error mechanisms in experimental platforms, our goal is to find a low-overhead Clifford-based representation for thermal relaxation error channels. Observing the decomposition of the amplitude damping error channel in Eq.~\eqref{eq:amp_decomp}, the negativity is contributed from the Pauli-Z operation branch. However, thermal relaxation typically includes additional dephasing $\gamma_\phi \neq 0$, which itself can naturally be modeled as a Pauli-Z error channel (see Sec.~\ref{sec:theory:thermal}). This suggests that, by creating an affine combination of the amplitude damping and dephasing error channels, the overall negativity of the thermal relaxation channel can be significantly reduced, yielding an efficient simulation method. 

%\textcolor{red}{Unify all your equations: some currently have ',' at the end, some with '.', some without anything.}
% Fixed.

\subsubsection{Zero temperature thermal relaxation}

We first consider the QPD of the thermal relaxation error channels at temperature zero, i.e., $\mathcal{E}_{\text{th}} = \mathcal{E}_{pd} \circ \mathcal{E}_{\text{ad}}$. The QPD of the thermal channel can be derived from Eq.~\eqref{eq:amp_decomp} and Eq.~\eqref{eq:dephasing} as
\begin{equation}
    \mathcal{E}_\text{th, 0} = q^{\text{(th, 0)}}_{+} \mathcal{I} + q^{\text{(th, 0)}}_{-} \mathcal{Z} + p_\gamma \mathcal{R}_{\ket{0}},
    \label{eq:th0_qpd}
\end{equation}
where the quasi-probabilities are
\begin{align}
    q^{\text{(th, 0)}}_{\pm} & = \frac{1-p_\gamma \pm (1 - 2 p_\phi)\sqrt{1- p_\gamma}}{2} \nonumber \\
    & = [e^{-\gamma \tau} \pm e^{-(\gamma/2 + \gamma_\phi)\tau} ] / 2 \nonumber \\
    & = [e^{-\tau/T_1} \pm e^{-\tau/T_2} ] / 2 . \label{eq:qth0}
\end{align}
From Eq.~\eqref{eq:qth0}, the decomposed identity channel from the thermal channel always has positive probabilities, while the Pauli-Z channel can have positive or negative quasi-probability, depending on the relative strength of the pure dephasing and energy relaxation process. Specifically, when $T_2 \leqslant T_1$, $q_\mathcal{Z} \geqslant 0$, implying that the combined thermal relaxation channel admits a fully positive decomposition in this regime, removing all QPD sampling overhead. Combining the amplitude damping with the dephasing, therefore, enables an exponential advantage over decomposing them separately. 

% which means there is a completely positive decomposition of the combined thermal relaxation channel. In this case, compositing relaxation with phase decoherence has an exponential advantage over sampling separate quasi-probability channels. 

% \begin{figure}[h]
%     \centering
%     \includegraphics[width=\linewidth]{figures/overhead_3D_composite.pdf}
%     \caption{Sampling overhead for the quasi-probability relaxation channel. $\Gamma \equiv 1$ in the composite method (the distribution is positive) when $T_1 = T_2$, and is greater than 1 everywhere with quasi-probability decomposition alone.}
%     \label{fig:sampling_overhead}
% \end{figure}

\begin{figure}[htbp]
    \centering
    \includegraphics[width=0.90\linewidth]{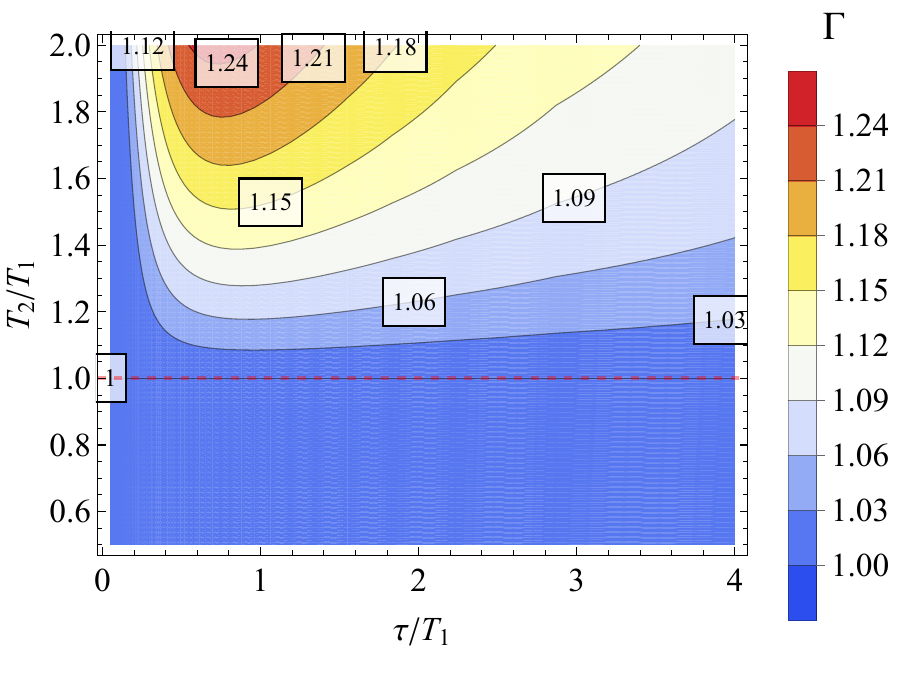}
    \caption{Sampling overhead for the quasi-probability relaxation channel on a single qubit, decomposed using Eq.~\eqref{eq:th0_qpd}. The overhead is measured by the negativity of the quasi-probabilistic distribution, which is captured by $\Gamma$. When $T_2/T_1 \leqslant 1$, the QP distribution is completely positive, while $T_2/T_1 = 2$, there is no additional dephasing, the overhead is the same as the QP distribution of an amplitude damping channel.}
    \label{fig:sampling_overhead}
\end{figure}

Importantly, this parameter regime is not only of theoretical interests, however, it is typical of many current quantum platforms, including superconducting and semiconductor spin qubits. In these systems, the environmental dephasing typically limits the coherence of the qubits, such that the $T_2$ time is comparable to $T_1$~\cite{shankar_spin_2010, rigetti_superconducting_2012}. Thus, many practically relevant quantum devices naturally fall into this ``sweet spot'' where the decomposition results in a fully positive distribution, costing no extra sampling overhead. Specifically, in the limiting case $T_1 = T_2$, the thermal relaxation channel further simplifies to a pure stochastic reset process.  

\begin{figure}[htbp]
    \centering
    \includegraphics[width=0.9\linewidth]{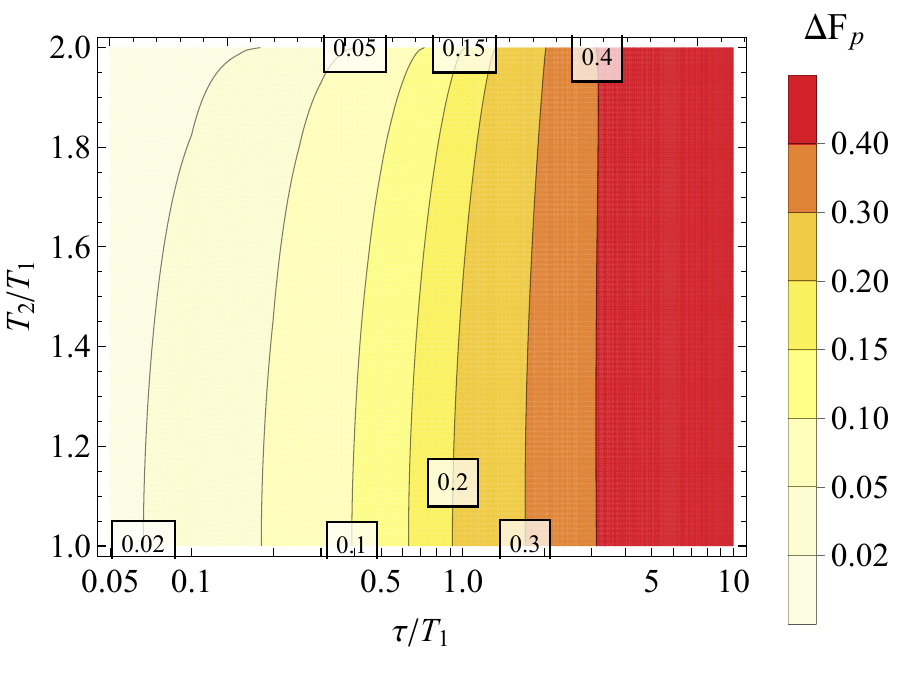}
    \caption{We compare the channel fidelity of the reset error and the Pauli-twirled thermal channel to the exact thermal channel at zero temperature. We plot the difference of the channel fidelity $\Delta F = F_{\text{reset}} - F_{\text{pw}}$. In this regime, the reset channel can always give a better channel fidelity to the exact thermal relaxation error channel compared to the Pauli-twirled error channel. }
    \label{fig:channel_fid_compare}
\end{figure}

When $T_1 < T_2$, the decomposition is no longer completely positive, but the affine combination can still be simulated exactly by sampling over the quasi-probabilities. In this case, combining phase decoherence with the amplitude damping helps to lower the overhead, as seen in Fig.~\ref{fig:sampling_overhead}. Once $T_2$ coherence is at the relaxation limit, i.e., there is effectively no pure phase decoherence, the composite method converges to the standalone relaxation stabilizer decomposition found in Ref.~\cite{bennink_unbiased_2017}. This is consistent with the findings in Ref.~\cite{hakkaku_sampling-based_2021}, where combining Pauli-X error channels with coherent errors can reduce the sampling overhead. 

Beyond exactly sampling from the QP distribution, in the $T_2 \gtrsim T_1$ regime, one may approximate the thermal relaxation channels by the reset error channel, trading accuracy for efficiency by maintaining the complete positivity of the distribution. 
% where one can trade simulation accuracy for efficiency with no overhead. More 
Specifically, we consider the quantum channel at absolute zero
\begin{equation}
    \mathcal{E}_\text{reset} = q_{+}^{\text{(th, 0)}} \mathcal{I} + (1-q_{+}^{\text{(th, 0)}}) \mathcal{R}_{\ket{0}},
    \label{eq:th0_reset}
\end{equation}
which matches the exact QPD thermal relaxation channel in Eq.~\eqref{eq:th0_qpd} when $T_2 = T_1$. When $T_2 > T_1$, this reset channel neglects the negative term $q_{-}^{\text{(th, 0)}}$, and the reset probability is re-scaled. 
% Comparing to the exact QPD of the thermal relaxation channel in Eq.~\eqref{eq:th0_qpd}, the negative quasi-probability term $q_{-}^{\text{(th, 0)}}$ is ignored, and the probability for the reset operation is rescaled. This quantum channel is completely positive, thus no additional sampling overhead. 

To better quantify the accuracy of this channel, we compute the process fidelity to the exact thermal relaxation error channel by 
\begin{equation}
    F_{\mathcal{E}} = F\left( (\mathcal{E}_{\text{th,0}}\otimes\mathcal{I})[\rho_\text{me}], (\mathcal{E}\otimes \mathcal{I})[\rho_\text{me}] \right),
\end{equation}
where $\rho_\text{me}$ is a maximally entangled state defined on two copies of the Hilbert space that the quantum channels $\mathcal{E}$ and $\mathcal{E}_{\text{th,0}}$ act on~\cite{mayer_quantum_2018}. In our case, the maximally entangled state is a GHZ state $\ket{00} + \ket{11}$. We compare it with the Pauli-twirled thermal relaxation channel. Fig.~\ref{fig:channel_fid_compare} shows the channel fidelity gain of $\mathcal{E}_\text{reset}$ with respect to the Pauli-twirled channel $\mathcal{E}_\text{th,pta}$ in Eq.~\eqref{eq:th0_pta}. We focus on the regime where no exact positive QPD exists, $2\geq T_2/T_1>1$. In the experimentally relevant regimes, the reset channel $\mathcal{E}_\text{reset}$ shows constantly better channel fidelity to the exact thermal relaxation channel. We notice that the difference between the two error channels decreases as $\tau/T_1$ decreases. This is because as we reduce the ratio $\tau/T_1$, the error probability decreases, which makes both the PTA and the reset channel closer to the identity channel, thus reducing the distance between the two channels. But the reset channel $\mathcal{E}_{\text{reset}}$ always has better channel fidelity to the exact thermal relaxation channel. As the thermal relaxation error becomes increasingly pronounced, the advantage of the reset channel becomes more significant. We attribute this to the fact that the reset error channel can efficiently capture the directional bias of the thermal relaxation error channel in the low-dephasing regime while the PTA channel fails (see Fig.~\ref{fig:bloch}).

% \begin{figure}[h]
%     \centering
%     \includegraphics[width=\linewidth]{figures/trace_distance.pdf}
%     \caption{Trace distance between $\ket{+}$ Density Matrix-calculated reference state, and 10,000 shot sampling approximations with the composite stabilizer decomposition and PTA. The composite method can be refined to be exact when $T_2/2\leq T_1<T_2$ by performing quasi-probability sampling. In this figure, for direct comparison with an equal number of samples, the quasi-probability function is naively re-normalized to be completely positive, then Monte-Carlo sampled.}
%     \label{fig:sampling_positive}
% \end{figure}

\subsubsection{Finite temperature thermal relaxation}

\begin{figure}[htbp]
    \centering
    \includegraphics[width=0.95 \linewidth]{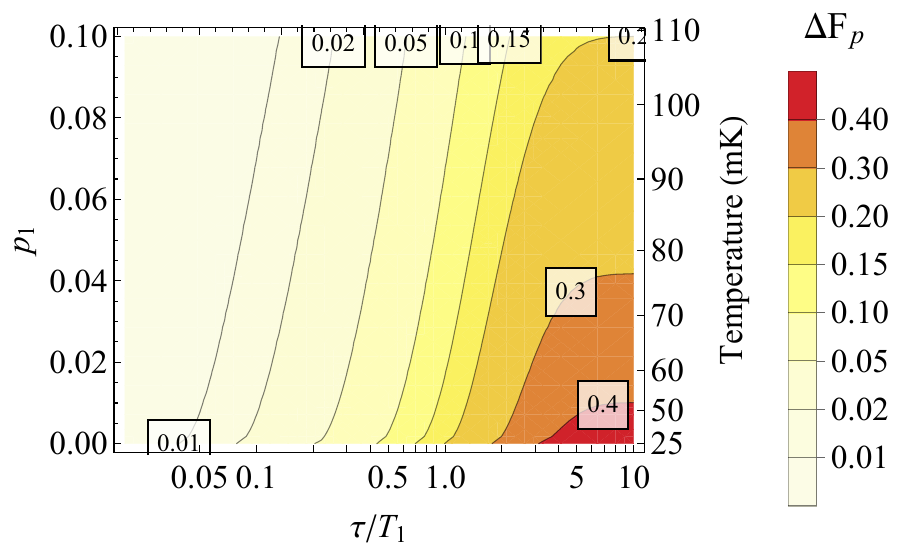}
    \caption{The channel fidelity gain of the reset-based Clifford channel to the PTA channel to approximate a finite temperature thermal relaxation channel. We set $T_2 = 1.5 T_1$. We sweep the excited state population up to $p_1 = 0.1$. To give a more intuitive understanding of experimental relevance, we compute the corresponding thermal temperature assuming the qubit frequency is $5$~GHz, which is labeled on the right.}
    \label{fig:finite_t_fid}
\end{figure}

We now extend the decomposition to the thermal relaxation channel at nonzero temperature. We start from the general amplitude damping channels, whose Kraus operators are Eqs.~\eqref{eq:gad_kraus_01},~\eqref{eq:gad_kraus_2}, and~\eqref{eq:gad_kraus_3}. As the Kraus operators $E_{\text{gad}, 0,1}$ only differ from the amplitude damping Kraus operators by only a constant scaling factor, this partial channel can be decomposed similarly as in Eq.~\eqref{eq:amp_decomp}. The branches that describe the thermal excitation $E_{\text{gad},2,3}$ can be viewed as an amplitude damping channel conjugated by Pauli-X, i.e., 
\begin{align}
    \mathcal{E}_\text{gad} = (1-p_1) \mathcal{E}_\text{ad}[p_\gamma] + p_1 \mathcal{X} \circ \mathcal{E}_{\text{ad}}[p_\gamma'] \circ \mathcal{X},
\end{align}
where $\mathcal{X}$ is the quantum channel for the Pauli-X, $p_\gamma$ and $p_\gamma'$ are given in Eq.~\eqref{eq:pgamma_temp}. Thus, the QPD of the general amplitude damping channel becomes
\begin{align}
    \mathcal{E}_\text{gad} = & q^{\text{(gad)}}_{+} \mathcal{I} +q^{\text{(gad)}}_{-} \mathcal{Z} \nonumber \\
    & + (1-p_1)p_\gamma \mathcal{R}_{\ket{0}} + p_1 p_\gamma \mathcal{R}_{\ket{1}},
\end{align}
where the quasi-probabilities are
\begin{align}
    q^{\text{(gad)}}_{\pm} & = (1-p_1) q^{\text{(ad)}}_{\pm}[p_\gamma] + p_1 q^{\text{(ad)}}_{\pm}[p_\gamma] = q^{\text{(ad)}}_{\pm}.
\end{align}

Including the dephasing effect into consideration, the full finite-temperature thermal relaxation channel becomes,
\begin{align}
    \mathcal{E}_\text{th} = & q^{\text{(th)}}_{+} \mathcal{I} +q^{\text{(th)}}_{-} \mathcal{Z} \nonumber \\
    & + (1-p_1)p_\gamma \mathcal{R}_{\ket{0}} + p_1 p_\gamma \mathcal{R}_{\ket{1}},
\end{align}
where the quasi-probabilities are
\begin{equation}
    q^{\text{(th)}}_{\pm} = q^{\text{(th, 0)}}_{\pm}[p_\gamma],
\end{equation}
$q^{\text{(th, 0)}}_{\pm}$ is given in Eq.~\eqref{eq:qth0}, the thermal relaxation and excitation probability $p_\gamma$ are defined in Eq.~\eqref{eq:pgamma_temp}. 
We notice that compared to the zero-temperature thermal relaxation channel, the negativity of the distribution remains the same as the zero-temperature thermal channel with the same $p_\gamma$, indicating that higher temperatures do not increase or decrease the sampling cost. In the high-temperature limit where $k_b T_b \gg \hbar \omega$, $\langle n_b\rangle \gg 1$, making the probability of the excited state $p_1 \sim 1/2$. In this regime, the thermal relaxation tends to thermalize towards the maximally mixed state $(\dyad{0} + \dyad{1})/2$ and the directional feature is lost. The thermal relaxation channel can be exactly represented by the PTA. However, for experimentally relevant thermal bath temperature $k_b T_b < \hbar \omega$, the reset error channel can still be a good approximation to the thermal relaxation error channel,
\begin{align}
    \mathcal{E}_\text{reset} =&  q_{+}^{\text{(th)}} \mathcal{I} + (1-q_{+}^{\text{(th)}}) (1-p_1)\mathcal{R}_{\ket{0}} \nonumber \\
    & + (1-q_{+}^{\text{(th)}}) p_1 \mathcal{R}_{\ket{1}}.
\end{align}

In Fig.~\ref{fig:finite_t_fid}, we plot the channel-fidelity improvement of the reset-based model relative to the PTA channel with finite temperature. We focus on the regime $p_1 \leqslant 0.1$, corresponding to $k_b T_b / \hbar \omega \sim 0.45$. For a $5$~GHz superconducting qubit, this corresponds to a bath temperature of approximately $T_b \sim 109$~mK. We take $T_2/T_1 = 1.5$ as an example. Across this regime, the reset model consistently achieves higher channel fidelity to the finite-temperature thermal-relaxation channel than the PTA channel. As the thermal temperature increases, however, the advantage of the reset model gradually diminishes, consistent with the analysis above.

\section{Numerical analysis of quantum memories under thermal error} \label{sec:sim}

% \sg{Addressing Chenxu's comments in the following paragraph}
Superconducting qubits are both well established and a promising avenue for scaled quantum computers~\cite{koch_charge-insensitive_2007}. Practical developments in multiple QEC directions on superconducting qubits have been made, including planar codes and quantum Low Density Parity Check (qLDPC) codes~\cite{acharya_quantum_2025, lacroix_scaling_2025, wang_demonstration_2025}. The rotated surface code~\cite{fowler_surface_2012} and Bicycle-bivariate (BB) qLDPC~\cite{yoder_tour_2025} are two particularly promising code variants towards fault tolerance. Rotated surface code offers an efficient planar layout with well-understood operators, while BB code offers a very high encoding rate with minimal layers of connectivity to support the sparsity necessary for qLDPC. Understanding the code performance under realistic noises, especially thermal relaxation, can greatly help to optimize the code design and support the experimental realization.

Prior studies have investigated small to medium surface code performance under thermal relaxation with PTA~\cite{ghosh_surface_2012, tomita_low-distance_2014}. For distance-3, comparisons with the exact thermal relaxation using density-matrix simulation were also performed~\cite{katabarwa_logical_2015, tomita_low-distance_2014, puzzuoli_tractable_2014}. PTA in these simulations was found to have a $2$-$10\times$ overestimation of error depending on the code and error parameters. Even distance-3 codes are difficult to fit within the density-matrix simulation. For previous studies, syndrome qubits were reused to ``truncate'' the code, which does not preserve the time order that physical surface codes are measured in. Surface codes larger than distance-3 have too many data qubits for exponential methods with or without truncation to be practical.

On the side of qLDPC codes, the noise performance of the BB code is thus-far under investigated. BB-codes with the best encoding rate are out of reach for density-matrix simulation. To our knowledge, the BB code logical error rate simulation has only been performed using uniform depolarizing errors to fit existing stabilizer simulators. BB codes beyond [[18, 4, 4]] are targeted for fault-tolerance and high encoding rates, but are far too large for simulating thermal noise with exponential methods. By using the decomposition model discussed in Sec.~\ref{sec:theory:QPD_thermal} to simulate $T_1$ and $T_2$ errors with stabilizer sampling, we are able to explore targeted qLDPC codes and preserve syndrome qubit layouts with both accuracy and efficiency.

\begin{table}[h!]
\centering
\caption{\label{table:IBM_data} The median coherent times ($T_1$ and $T_2$) of IBM Heron QPUs~\cite{ibm_quantum_computers}.}
\begin{ruledtabular}
\begin{tabular}{lccc}
\textbf{Device} & $\mathbf{T_1} \  (\mu\text{s})$ & $\mathbf{T_2} \ (\mu\text{s})$ & $\mathbf{T_1/T_2}$ \\
\hline
Boston & 275.27 & 338.82 & 0.81\\
Fez & 142.41 & 98.43 &  1.45\\
Kingston & 261.36 & 131.93 & 1.98\\
Marrakesh & 185.77 & 104.16 &  1.78\\
Pittsburgh & 300 & 324.17 & 0.93 \\
Torino & 183.67 & 131.48 & 1.40\\
\hline
\textbf{Average} & \textbf{242.75} & \textbf{188.165} & \textbf{1.29} \\
\end{tabular}
\end{ruledtabular}
\end{table}

It happens that transmon qubit technology falls within $T_1\approx T_2$, and thus can be simulated under composite stabilizer decomposition with good efficiency and accuracy. Table~\ref{table:IBM_data} provides coherence times at the time of writing for the current generation Heron processors, which vary with calibrations and individual operations. Current transmon-based hardware has $T2 \sim T1$ with fluctuations. In our numerical investigation of the surface and BB codes, we assume $T_1 = T_2$ for simplicity. So long as $T_1\geq T_2$, the composite method will have no overhead for exact simulation.

In the rest of this section, we adopt the thermal relaxation error model developed in Sec.~\ref{sec:theory:QPD_thermal}, and integrate it into our highly efficient Clifford simulator. Sec.~\ref{sec:sim:simulator} introduces the simulator and experimental settings. The surface code and BB code logical performances are reported in Sec.~\ref{sec:sim:surface} and~\ref{sec:sim:bb}, respectively.

% \sg{Add that BB code hasn't been simulated exactly before. Also, previous surface code studies have removed syndrome qubits to fit DM sim. This gets rid of time order capabilities.}

\subsection{Clifford Simulator} \label{sec:sim:simulator}

% \cl{In the first half of this subsection, we need to discuss the features we included in the simulator and the features from stim that cannot support the efficient simulation of the qec code with quasi-probabilities. Some key points include why stim cannot easily simulate the noise model we discussed, why pauli frame tracking in stim does not naturally fit, what we have done to address this (the feature our stabilizer has), why we need gpu, etc. But basically, we need the discussion of the simulator. You can think of this as a very concise version of your stab-sim paper.}
% \sg{working on this now}

To perform these simulations, we use a new stabilizer simulator~\cite{garner_stabsim_2025} with MPI and GPU support for many-shot simulation and large-code speedup, respectively. Like other state-of-the-art simulators, including Stim~\cite{gidney_stim_2021}, our simulator, STABSim~\cite{garner_stabsim_2025}, is derived from the stabilizer and destabilizer tableaus introduced in CHP~\cite{aaronson_improved_2004}. Thermal relaxation and decoherence errors are modeled as probabilistic gates injected into the circuit during large timesteps or idles. Every error gate performs its own Monte-Carlo roll via random number generation to choose between potential Clifford or Reset operations based on the stabilizer decomposition probability distribution. For $T_1 \geq T_2$, the thermal relaxation can be exactly decomposed into a probability distribution, inducing no sampling overhead. When $T_1 < T_2$, the negativity in the QP distribution can be handled in two ways: (i) the QP distribution can be sampled exactly with a smaller exponential number of shots by renormalizating the QP distribution to the corresponding probability distribution shown as in Eq.~\eqref{eq:unbiased_estimator}, or (ii) the thermal relaxation channel can be approximated by the reset channel introduced in Eq.~\eqref{eq:th0_reset}, and sample it using Monte-Carlo methods when the sampling cost is a concern.
%, the quasiprobability function can be naively renormalized and Monte-Carlo sampled as if the function was always positive. For $T_1$, $T_2$ damping, this remains more accurate than PTA~\ref{fig:sampling_positive}. Otherwise, the function can be sampled with exponential shots with quasiprobability renormalization as in Eq.~\ref{eq:unbiased_estimator} for exactness. 

Our new framework provides a few standout differences for QEC simulations versus Stim~\cite{gidney_stim_2021}: (i) Stim is optimized for distributing single-thread CPU performance, however, large stabilizer tableaus can be greatly sped up with GPU parallelization, and (ii) the Pauli frame tracking that underlies Stim's fast sampling in noisy simulations cannot admit probabilistic reset operations, limiting it to Pauli-Twirled approximations only. For the numerical simulations here, we distribute shots with MPI across CPU threads to probe lower error rates. Our CPU simulator provides faster performance in the tens to low-hundreds of qubit range that single logical qubit codes will use~\cite{garner_stabsim_2025}. 

% Furthermore, superconducting qubits tend to be $T_1$-dominated in their error~\cite{}, so improving relaxation error modeling is significant in understanding which codes and at what coherence times algorithms can be practically considered.

\begin{figure*}[htpb]
    \centering
    \subfloat[Logical $\ket{0}_L$]{
    \includegraphics[width=0.37 \linewidth]{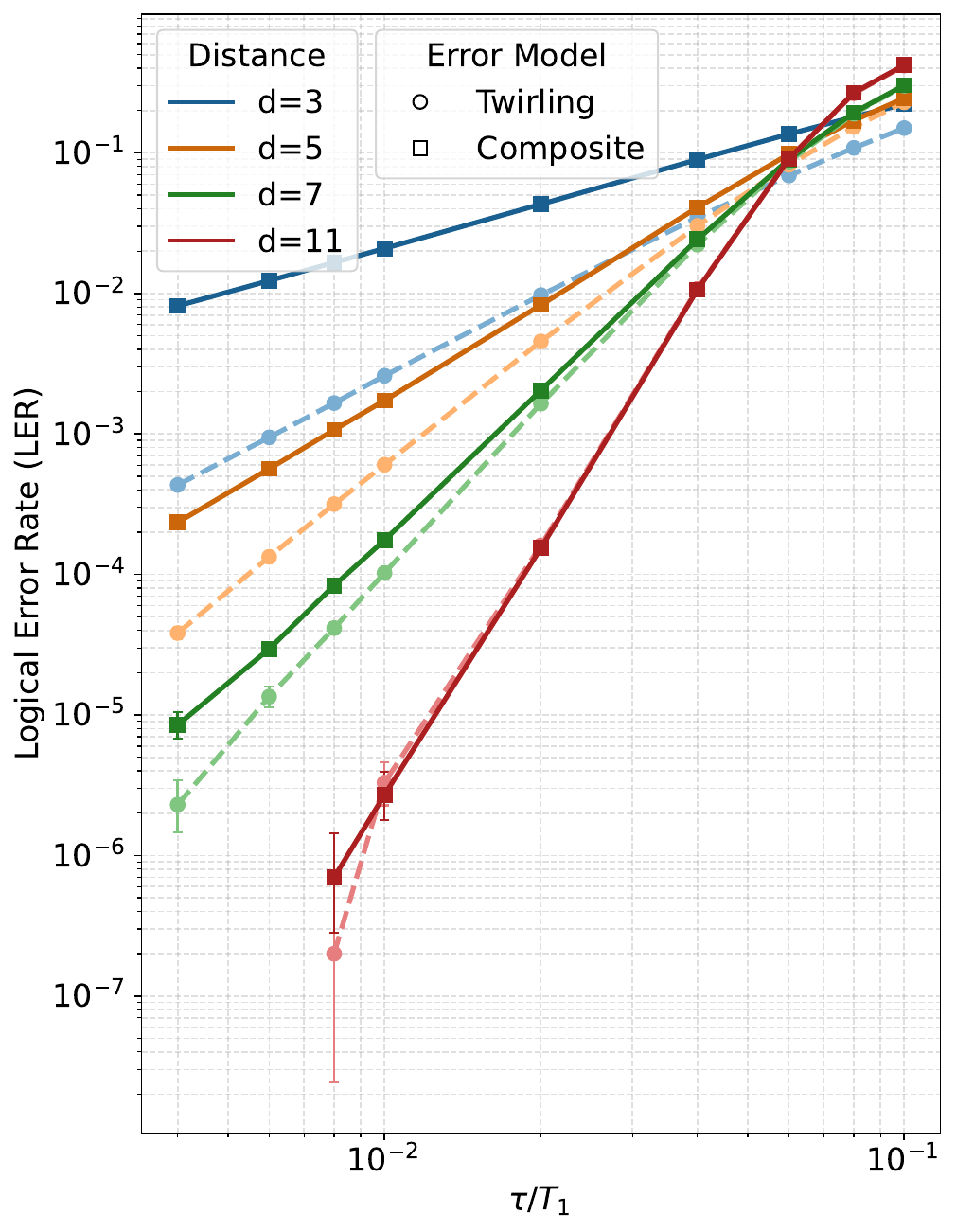}
    } \quad
    \subfloat[Logical $\ket{+}_L$]{
    \includegraphics[width=0.37 \linewidth]{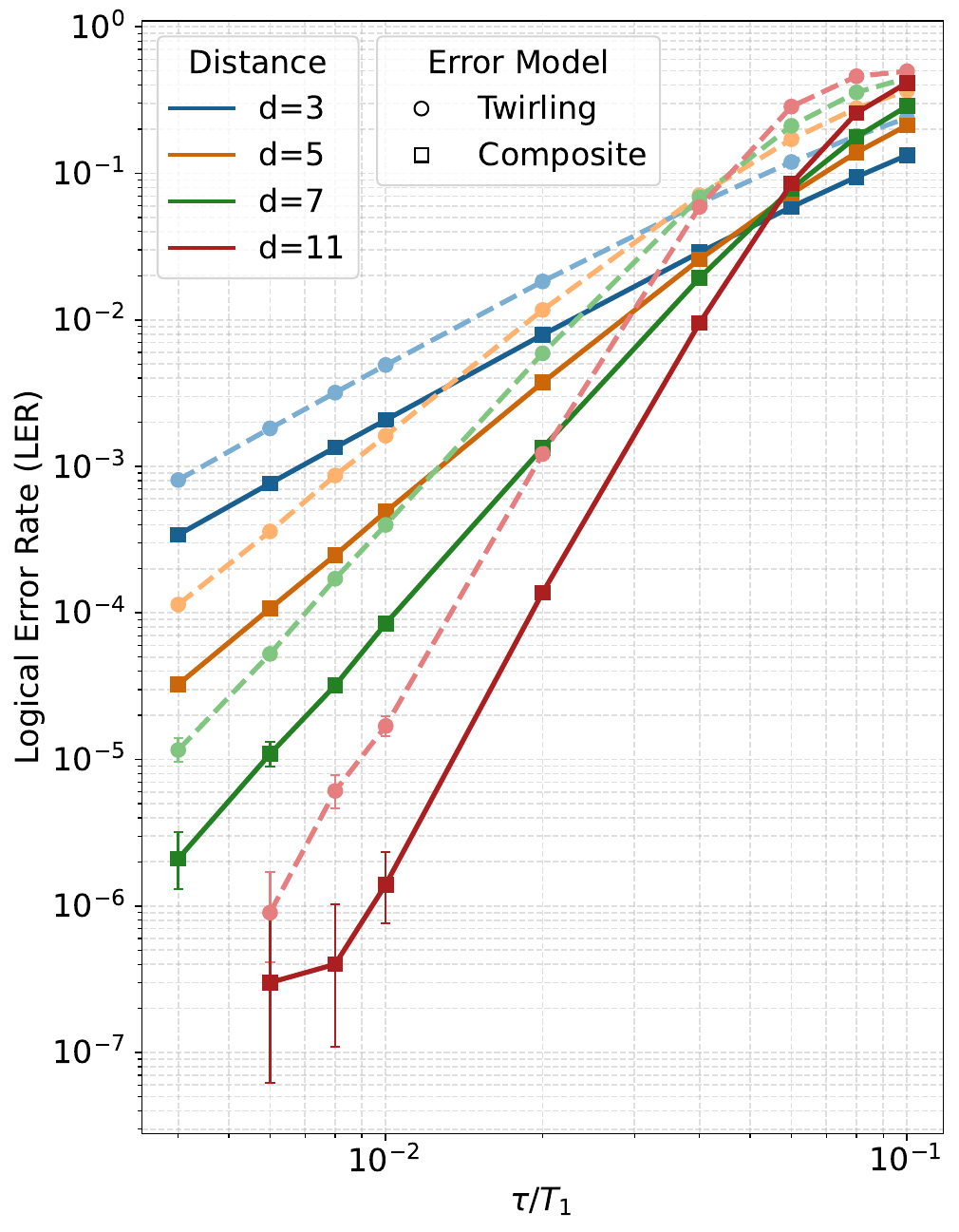}
    }
        \caption{Rotated Surface Code $\ket{0}_L$ (a) and $\ket{+}_L$ (b) memory experiments decoded using Minimum Weight Perfect Matching, $10^7$ shots per data point. We only track the thermal relaxation errors during the syndrome qubit measurement, whose time ($\tau$) is scaled by the device $T_1$. We calculate the LER of the surface code with the exact thermal relaxation error channel (dashed lines) and the channel under PTA (dashed lines). Error bars show a 95\% confidence interval for the number of logical errors given 10 million shots}
    \label{fig:surface_code}
\end{figure*}

\subsection{Rotated Surface Code} \label{sec:sim:surface}

Rotated surface code fits the design constraints of local connectivity for planar superconducting hardware architecture while saturating the BPT bound in two dimensions~\cite{bravyi_tradeoffs_2010}. It uses fewer resources than the planar surface code, and supports universal logical operations via state injection and lattice surgery~\cite{horsman_surface_2012}. To understand how a rotated surface code would perform as memory under exact thermal relaxation, we apply the thermal relaxation error channels before measurement operations. This is because the measurements are typically much slower on superconducting devices than one and two-qubit gates, and it is the most heavily affected by the thermal relaxation errors. We specifically focus on thermal channels under PTA and the exact channel with the error model discussed in Sec.~\ref{sec:theory:thermal}. This application of noise focuses on the incoherent effects of relaxation and decoherence, complementing recent work on better modeling for coherent gate noise~\cite{hakkaku_sampling-based_2021}.

In Fig.~\ref{fig:surface_code}, we investigate rotated surface code memories, with particular interest in larger distance codes to understand how performance scales under our exact model. We simulate the logical performance of surface code with distance $d = 3$, $5$, $7$, and $11$. In each memory experiment, we run $d$ rounds of syndrome checks, and use the minimum-weight-perfect-matching (MWPM) decoder to decode the syndromes and solve for the logical error rates (LER). The LER under thermal relaxation when the code block is in logical $\ket{0}_L$ (Fig.~\ref{fig:surface_code}a) and logical $\ket{+}_L$ state (Fig.~\ref{fig:surface_code}b) are computed. Since measurement dominates the idling time in superconducting systems, thermal errors are injected on all qubits while they await measurement extraction on syndrome qubits. We observe that PTA channels often predict the code performance reasonably with a constant offset, consistent with the previous study~\cite{katabarwa_logical_2015}. However, with access to low error rate and large distance regimes, we find scaling is not perfect. When the logical state is $\ket{0}_L$, the PTA channel underestimates the LER, while in the state $\ket{+}_L$, the PTA overestimates. 

By undergoing a non-uniform channel like thermal relaxation, codes will perform differently based on the logical state stored. On hardware, classical Pauli frame tracking and other runtime techniques will be used to minimize unnecessary physical Pauli operations and make the quantum computation as effective as possible. The rotated surface code experiments shown in Fig.~\ref{fig:surface_code} act differently with the same damping models, as the data qubits initialized in Z and X can accumulate errors for many rounds of syndrome measurement without being reset. In low-distance Z-memory experiments, where the overall error probability is low, random bit flips on the data qubit are more easily decoded since our MWPM decoder cannot account for the state precession towards $\ket{0}$ in its priors. With the data qubits initialized in a $\ket{+}$ state for X-memory, decoder performance is more consistent with error rate and distance. In this case, noise bias from damping is only one component of undetected logical errors in data qubits. Probabilistic phase flips that arise from pure phase decoherence can be accounted for in MWPM priors such that the LER from PTA and stabilizer decomposition channels match.

% \begin{figure}[h]
%     \centering
%     \includegraphics[width=\linewidth]{figures/surfacecode_ler_x.pdf}
%         \caption{Rotated Surface Code $\ket{+}_L$ experiments decoded using Minimum Weight Perfect Matching, 10M shots per point. The number of syndrome extraction rounds equals code distance. Relaxation and decoherence errors are placed before measurement in each round, while one and two qubit gate times are taken to be much shorter in $\tau$ than measurement. Error bars show a 95\% confidence interval given the logical error rate and number of samples.}
%     \label{fig:surface_code_x}
% \end{figure}

\begin{figure}[htbp]
    \centering
    \includegraphics[width=0.8\linewidth]{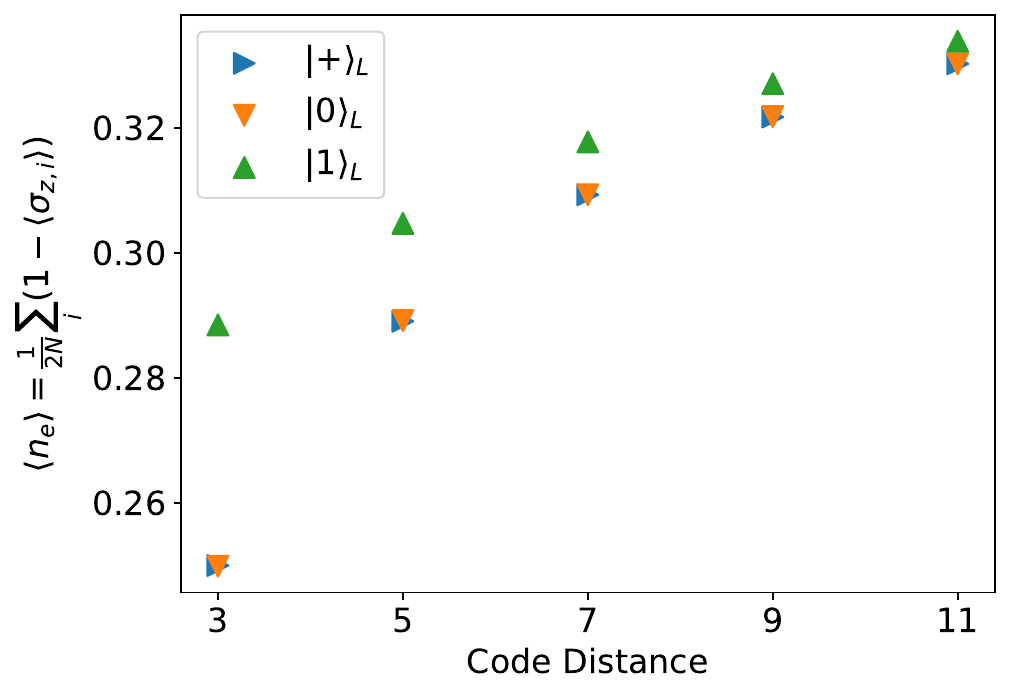}
        \caption{Excited state population of the rotated surface code block when the code is in logical $\ket{0}_L$, $\ket{1}_L$, or $\ket{+}_L$ state. We include both the data and syndrome qubits after the logical state preparation circuit for each logical state. As we did not post-select from the syndrome check outcomes, we are effectively averaging all the possible physical realizations of the same logical state. The generation circuits are shown in Appendix~\ref{app:circ}. Each point uses 10 million samples.}
    \label{fig:surface_code_exc}
\end{figure}

In Fig.~\ref{fig:surface_code_exc}, we plot the average excitation of the surface code block when the code block is in logical state $\ket{0}_L$, $\ket{1}_L$, and $\ket{+}_L$. We noticed that the excited state population of the logical $\ket{0}_L$ and $\ket{+}_L$ states differs only slightly, indicating that their LER differences are driven primarily by the MWPM decoder's response to the noise bias, rather than by the small differences in excited state population. A distance $d$ rotated surface code requires a minimum of $d$ qubits to support a logical operator, meaning a logical $\ket{1}_L$ state memory can be constructed such that the typical excited state population is very close to the logical $\ket{0}_L$ state. In testing LER curves for both error models in the $\ket{1}_L$ state are nearly identical to the curves in the $\ket{0}_L$ state (see Appendix~\ref{app:logical1}). The difference in scaling between memory states in Fig.~\ref{fig:surface_code} suggests that noise bias-aware decoding will be necessary in codes that can be solved with generalizations of MWPM. %\cl{Not quite sure about the meaning of the last sentence. May need revision.}\sg{fixed the wording}

% Since total excited state populations only vary slightly for a minimal logical operator, the effect on logical error rates relative to the noise bias impact on MWPM is small. 

\subsection{Bivariate Bicycle Code} \label{sec:sim:bb}

\begin{figure}[htbp]
    \centering
    \includegraphics[width=0.9 \linewidth]{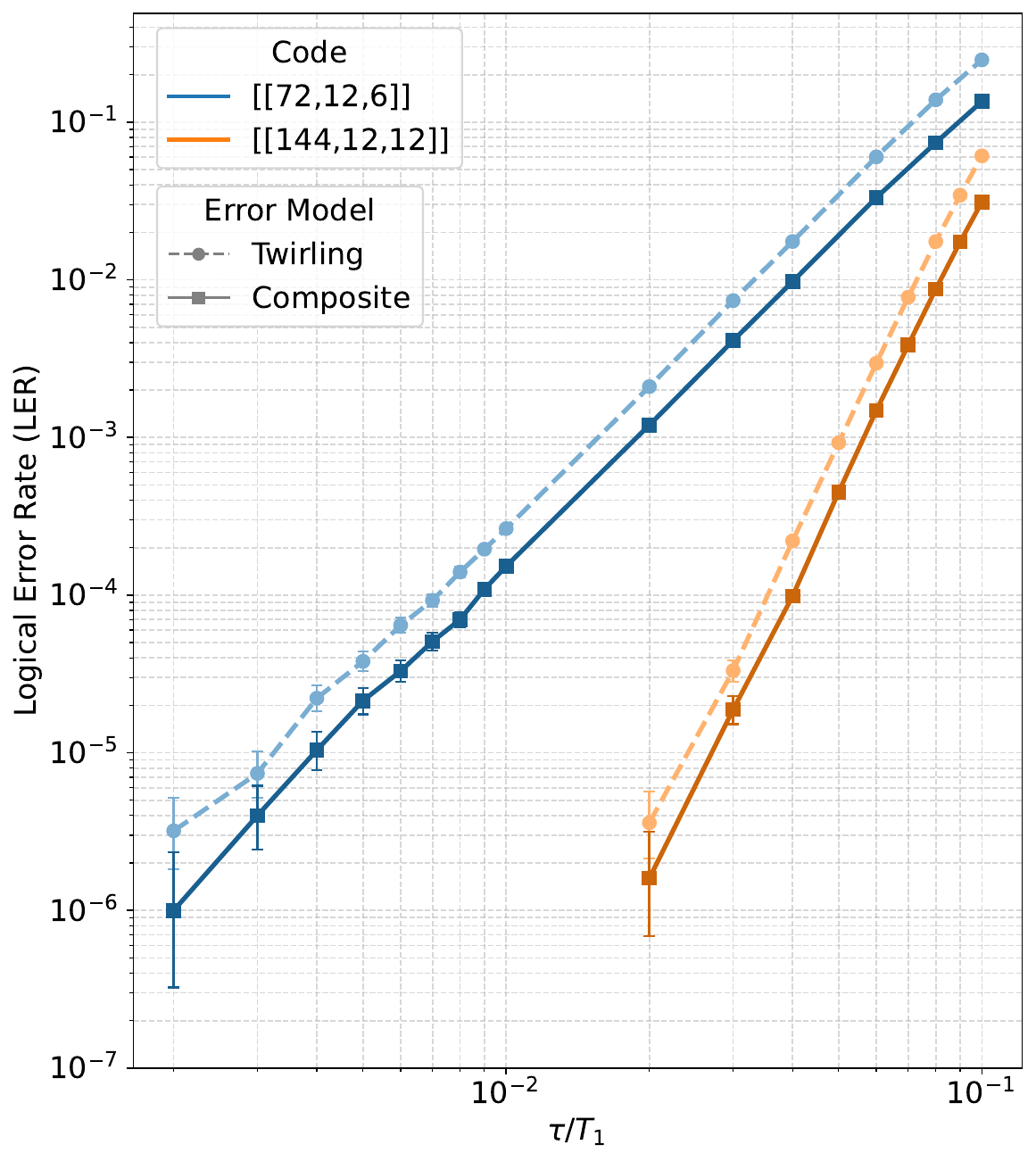}
    \caption{The LER of Bicycle Bivariate code memory experiments decoded with BP-OSD decoder. During each round of syndrome extraction, measurement resets are preceded and followed by thermal relaxation channels, with measurement duration $\tau$. LERs obtained using the exact thermal relaxation are shown as solid lines, while those using the PTA channels are plotted as dashed lines. Error bars show a 95\% confidence interval given 5 million samples.}
    \label{fig:bb_code}
\end{figure}

Bivariate Bicycle (BB) Code is a high encoding rate qLDPC scheme with minimal overlapping connection layers, helpful for superconducting architectures. BB code is not yet known to support universal quantum operations on its own, however, as a memory, it is capable of suppressing logical errors very effectively with small improvements in physical error rate, creating an efficient fault-tolerant universal architecture via lattice surgery with other codes~\cite{bravyi_high-threshold_2024, stein_hetec_2025}. Previous studies developing and introducing these codes have primarily used uniform depolarizing errors~\cite{yoder_tour_2025} for simplicity and simulation performance. 

Sparsely connected codes like BB code cannot be decoded with the same minimum weight matching graph method commonly employed for surface code and other locally connected codes. Instead, the state-of-the-art decoding algorithm for BB codes is Belief Propagation (BP), which propagates informed messages along the edges of the parity check Tanner graph~\cite{poulin_iterative_2008}. BP decoding is often followed by Ordered Statistics Decoding (OSD) refinement, which refines the BP output by flipping the least reliable bits and checking candidate error patterns at a combinatorial time cost~\cite{kung_belief_2023}. In our study, we use BP-OSD with up to 1000 BP iterations and 10 OSD iterations as a common baseline decoder to compare the composite thermal relaxation error model against the PTA.

% unreliable bits are flipped and checked at a combinatorial time cost~\cite{}. We choose BP-OSD with a maximum of 1000 BP iterations and 10 OSD iterations as a common baseline decoder to test the composite noise model against the PTA. 

In Fig.~\ref{fig:bb_code}, we investigate the impact of the thermal relaxation error on the high-encoding-rate BB code memories. Similar to the surface code experiments, we perform $d$ rounds of syndrome extraction for distance $d$ codes. Thermal relaxation errors are injected before each measurement round to all the qubits. The resulting syndromes are processed using the BP-OSD decoder~\cite{roffe_decoding_2020, Roffe_LDPC_Python_tools_2022} to numerically compute the LER. The PTA thermal relaxation channel shown in Fig.~\ref{fig:bb_code} serves as a baseline for real thermalization performance, while previous studies use uniform depolarizing errors without a direct connection to noise. The decomposed channel shows an offset from PTA, with approximately $1.5$-$2\times$ lower error rate relative to PTA, consistent with the trends observed in surface code experiments.

In BB codes, different logical qubits can share subsets of data qubits, determined by the generating polynomials of the code. This code structure disperses errors across the physical qubits, reducing the likelihood that a small number of localized faults produce a logical failure, but at the cost of increased connectivity and a more demanding decoding procedure that benefits from numerical refinement. 
% Different logical qubits one BB code will use some of the same data qubits, defined by the polynomial that generates the code. This has the effect of dispersing errors across the physical layout of the code, providing the advantage of making a few localized errors much less likely to cause a fault, with the cost of extra connectivity and decoding work with numerical refinement. 
For the purpose of error modeling, we find that this disbursement aggregates to a nearly constant offset between the LER of the PTA channel and the exact thermal relaxation error channel, similar to what was found in earlier low-distance surface code studies. OSD refinement can further smooth out logical error behavior by correcting unreliable bits, e.g., a qubit that has decayed and no longer has the expected behavior in future parity checks. Additionally, when comparing the LER of the BB code with that of the surface code under realistic thermal-relaxation noise, we observe that the BB code can achieve comparable or even lower logical error rates while encoding substantially more logical qubits, highlighting the advantages provided by its high encoding rate.

\section{Discussion} \label{sec:discussion}

% \cl{
% some points we may need to discuss here:
% \begin{enumerate}[noitemsep]
%     \item supporing the qubit-wise thermal noise model to perform more accurate modeling.
%     \item The channel negativity related to the magic of the channel (cite the Fijii's paper (if I remembered this right). composing different error channels can potentially change the channel magic, extending the capability of the clifford simulators. 
%     \item combinig with the extended stabilizer simulation methods, or combining tensor-networks, this opens up an oppotunity for simulation frameworks with larger capability
%     \item The PTA or the reset channel be used to pre-train the the decoder to make it working better with QEC codes, improving the code performance with thermal errors.
%     \item On the theory side, connecting the channel fidelity/distance to the QEC code logical properties, like the error threshold, still requires further investigation.
%     \item rare event simulation
% \end{enumerate} 
% }

% \sg{support qubit wise T1 T2}

This work demonstrates that combining amplitude damping and dephasing into a unified thermal relaxation channel can substantially reduce or even eliminate the QP sampling overhead required for exact simulation of non-Clifford noises. This finding opens several interesting avenues for further research in both theoretical understanding and large-scale simulation of realistic quantum error correction codes.

A natural extension of this work is to incorporate non-uniform, qubit-wise thermal noise models into the stabilizer simulation framework for better characterizing the hardware performance in QEC code implementations. Real devices exhibit spatial and temporal varying coherence properties. An efficient stabilizer simulator that can incorporate hardware-specific, qubit-wise $T_1$ and $T_2$ values from device calibration data and apply distinct thermal relaxation errors to each qubit would enable substantially more accurate predictions of the QEC code performance, especially the logical error rates. 
%This feature will be incorporated into the next version of our stabilizer simulator. 

Our findings also connect to the broader questions of the robustness and magic of the quantum channels. Recent works by Hakkaku {\it et al.}, building on the robustness of the quantum channels~\cite{howard_application_2017,seddon_quantifying_2019}, showed that compositing coherent error channels and Pauli error channels can reduce the channel robustness and the overhead for simulation, which our work is consistent with~\cite{hakkaku_sampling-based_2021}. This opens up a question: what other useful channels, previously thought of as too sophisticated for stabilizer simulation, can these principles be applied to? While we focus on relaxation and decoherence for this work, there are other physical processes that continuously affect qubit operation, and can potentially be efficiently simulated by combining with other operations. This requires a systematic framework for identifying and optimizing the combinations to expand the capability of classical simulators. Furthermore, it is possible that commonly reused non-Clifford gate blocks, which contain a lot of magic can be combined in such a way that makes a minimally negative non-Clifford channel. If these channels can be combined with little negativity, there could be new shortcuts in algorithm design, or at least a rethink of the most critical advantages that quantum computing brings us. 

On the applications side, the approximate thermal-relaxation models developed here, such as the reset-based approximation, may also prove valuable for decoder design, training, and calibration. Recent development of machine-learning-based decoders shows great advantages in decoding speed and scalability to larger systems~\cite{ataides_neural_2025,bausch_learning_2024}. Using PTA or reset channels as pre-training models could help adapt decoders to the biased noise channels from the thermal relaxation and to hardware realities. Such noise modeling may improve logical performance in thermal-noise regimes.

From a theoretical perspective, the connection between channel-level metrics (e.g., channel fidelity, state distance) and logical-level properties (e.g., error threshold, bias sensitivity, or error suppression rate) remains incompletely understood. Establishing quantitative relationships between microscopic channel structure and macroscopic fault-tolerant behavior would provide a principled foundation for noise-tailored code design and decoder optimization.

Recent ongoing efforts in rare event simulation may also be helpful in combination with more accurate error modeling to investigate fault-tolerant designs. Higher-distance codes presented in Figs.~\ref{fig:surface_code} and~\ref{fig:bb_code} require tens of millions or more shots to encounter a logical error, even given the relaxation and measurement times of only current superconducting qubits, which are still improving. Many-shot fast stabilizer simulation and distribution via Sinter and Stim~\cite{gidney_stim_2021}, has played a significant role in enabling works up to this point. However, for large fault-tolerant operations including lattice surgery between logical qubits, works in recruiting parallelization for single-shot speedup~\cite{garner_stabsim_2025} and circuit-level rare event simulation~\cite{mayer_rare_2025} are promising for the viability of future quantum error correction work.

\section{Conclusion} \label{sec:summary}

In this work, we developed a stabilizer-compatible method for efficiently simulating the thermal relaxation error channels by combining amplitude damping and dephasing into a unified channel. We showed that in the experimentally relevant regimes $T_2 \leqslant T_1$, this composite error channel admits a completely positive Clifford decomposition with no quasi-probability sampling overhead. This decomposition enabled efficient and scalable simulation of realistic thermal relaxation processes within the stabilizer frameworks for QEC study and beyond. 

Using this approach, we integrate the thermal relaxation error channels into our GPU-accelerated stabilizer simulator and perform large-scale simulations of QEC codes to understand their logical performance under realistic noise conditions. We focused on superconducting qubit platforms where the qubits' $T_1$ and $T_2$ times fall within the sweet spot of the Clifford decomposition. We investigated the logical performance of the two most promising QEC codes in superconducting platforms, i.e., surface codes and the BB code, and compared their noisy performance with the channels under PTA. Across memory experiments for both code examples, we observe that PTA can misestimate logical error rates by 2-10x in either direction, depending on code distance and code type. We also observed that the rate of error suppression with exact thermal relaxation could depend on the logical states of the code block, which we attribute to the decoder given the closeness in excited state population. 

In the regime where $T_2 > T_1$, we further introduced a reset-based approximation to the thermal relaxation error channel. In this regime, we have shown that this reset approximation can consistently outperform the PTA while maintaining the complete positivity of the decomposition and compatibility with the stabilizer simulation framework. This model provides a practical route for realistic noise simulations without the cost of the full sampling overhead of QPD. 

These results establish a practical route for incorporating physically realistic thermal relaxation noise into fast stabilizer-based simulation frameworks. Approximations that preserve Clifford compatibility while more accurately capturing the intrinsic bias and directional structure of thermal relaxation noise can play an important role in designing and assessing quantum error correcting codes for fault-tolerant quantum computing.

% In conclusion, these results establish a pathway for incorporating physically realistic thermal relaxation noise models into fast stabilizer simulation frameworks. Noise models with better approximation to the biased noise feature while maintaining the Clifford compatibility can be impactful for improving the design and evaluating the performance of QEC codes for fault-tolerant quantum computing.

% We find for our memory experiments that a more accurate noise model can significantly affect the offset and slope of the error suppression curves. \cl{This paragraph sounds like a conclusion, should appear somewhere later.}
% \sg{Moved here for now. Probably could use some more rewording.}

% The findings presented here are helpful for guiding code and decoder development with a more accurate qubit thermalization. They also present an opportunity for large, simulated data sets as a ground truth for machine learning decoders. ML decoding is ongoing consideration alongside analytical decoders such as MWPM since it has the ability to parametrize biases not considered in equal weight matching algorithms~\cite{bausch_learning_2024, lacroix_scaling_2025}. 

\section*{Acknowledgments}
We thank Shifan Xu, Ming Wang, and Shuwen Kan for fruitful discussions on BB code and surface code decoding. The research on noise modeling was supported by Pacific Northwest National Laboratory’s Quantum Algorithms and Architecture for Domain Science (QuAADS) Laboratory Directed Research and Development (LDRD) Initiative.
The Clifford simulation work was supported by the U.S. Department of Energy, Office of Science, National Quantum Information Science Research Centers, Quantum Science Center (QSC).
N. M. M. was supported by the “Embedding QC into Many-body Frameworks for Strongly Correlated Molecular and Materials Systems” project, which is funded by the U. S. Department of Energy, Office of Science, Office of Basic Energy Sciences (BES), the Division of Chemical Sciences, Geosciences, and Biosciences under FWP 72689.
This research used resources of the Oak Ridge Leadership Computing Facility, which is a
DOE Office of Science User Facility supported under Contract DE-AC05-00OR22725. 
This research used resources of the National Energy Research Scientific Computing Center (NERSC), a U.S. Department of Energy Office of Science User Facility located at Lawrence Berkeley National Laboratory, operated under Contract No. DE-AC02-05CH11231.
The Pacific Northwest National Laboratory is operated by Battelle for the U.S. Department of Energy under Contract DE-AC05-76RL01830.

\bibliography{main-v1}

\appendix

\section{LER of surface code logical $\ket{0}_L$ and $\ket{1}_L$ states} \label{app:logical1}

\begin{figure}[htbp]
    \centering
    \includegraphics[width=0.8\linewidth]{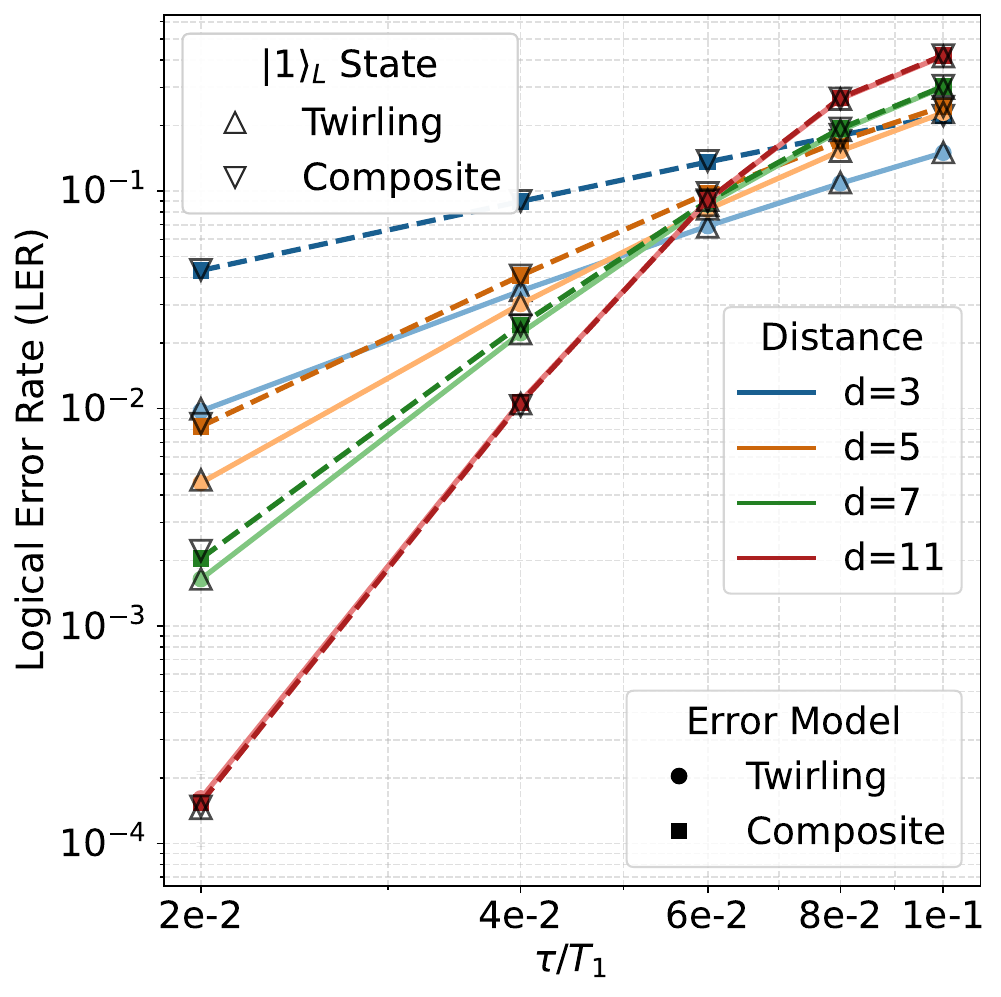}
    \caption{The logical error rate of the surface code memory experiments when the code is in logical $\ket{0}_L$ (solid colors) and logical $\ket{1}_L$ states (hollow markers).}
    \label{fig:logical1}
\end{figure}

In Fig.~\ref{fig:logical1}, we show the comparison of the LERs of the surface code memory experiments when the code is in state $\ket{1}_L$ and $\ket{0}_L$. We observe that the performance of the logical $\ket{1}_L$ state is nearly identical to the logical $\ket{0}_L$ state.

\section{Code Block Occupation Circuit} \label{app:circ}

\begin{figure}[htbp!]
    % \centering
    \includegraphics[width=\linewidth]{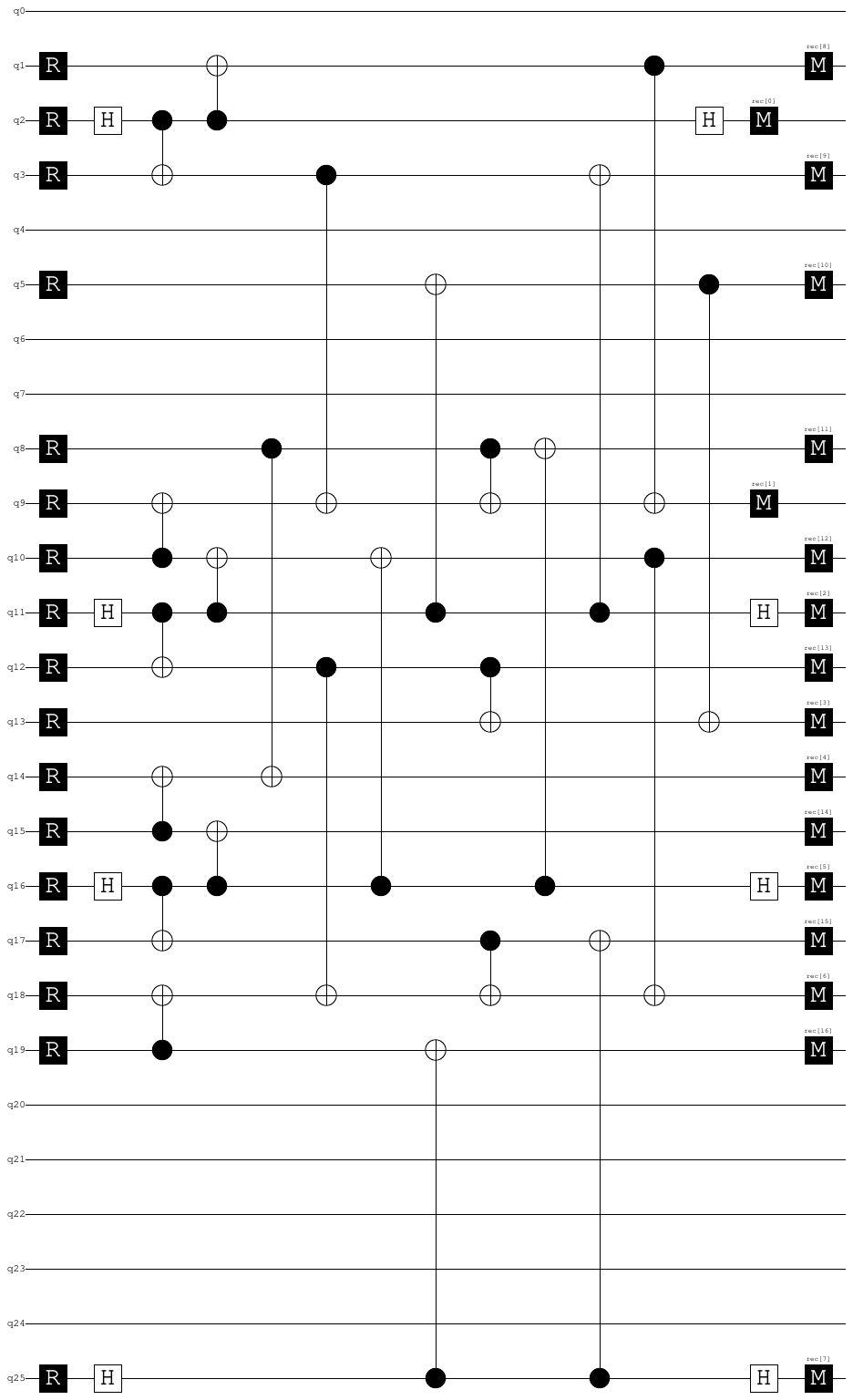}
    \caption{Logical state preparation circuit for state $\ket{0}_L$.}
    \label{fig:occupation_circuit}
\end{figure}

In Fig.~\ref{fig:occupation_circuit}, we show the circuit we use to compute the average occupation of the code block when the logical state is $\ket{0}_L$. We initially prepare all the physical qubits in state $\ket{0}$, then perform one round of syndrome checks. At the end of the circuit, we perform Pauli-Z measurements on all the qubits, including the data and syndrome qubits, to compute the average excitation probability. We do not perform post-selection on the outcomes of the syndrome qubits, and hence it is an average of all the possible physical realizations of the logical state $\ket{0}_L$. The other two logical states $\ket{+}_L$ and $\ket{1}_L$ are prepared similarly.

\end{document}